\documentclass[%
 reprint,
 amsmath,amssymb,
 aps,
]{revtex4-1}

\usepackage[dvipdfmx]{graphicx} 
\usepackage{dcolumn}
\usepackage{bm}
\usepackage[font=small,format=plain,labelformat=simple,labelsep=period,justification=raggedright]{caption}
\usepackage{amsmath}
\usepackage{here}
\usepackage{natbib}
\usepackage[dvipdfmx]{color}
\usepackage{epstopdf}

\begin{document}

\preprint{APS/123-QED}

\title{Dependence of the transportation time on the sequence in which particles with different hopping probabilities enter a lattice}

\author{Hiroki Yamamoto$^{1}$}
\email{h18m1140@hirosaki-u.ac.jp}
\thanks{}%
\author{Daichi Yanagisawa$^{2,3}$}%
\author{Katsuhiro Nishinari$^{2,3}$}%
\affiliation{%
$^{1}$ School of Medicine, Hirosaki University,
5 Zaifu-cho Hirosaki city, Aomori, 036-8562, Japan\\
$^{2}$ Research Center for Advanced Science and Technology, The University of Tokyo,\\
4-6-1 Komaba, Meguro-ku, Tokyo 153-8904, Japan\\
$^{3}$ Department of Aeronautics and Astronautics, School of Engineering, The University of Tokyo,\\
7-3-1 Hongo, Bunkyo-ku, Tokyo 113-8656, Japan
}%

\date{\today}

\begin{abstract}
\noindent
Smooth transportation has drawn the attention of many researchers and practitioners in several fields. In the present study, we propose a modified model of a totally asymmetric simple exclusion process (TASEP), which includes multiple species of particles and takes into account the sequence in which the particles enter a lattice. We investigate the dependence of the transportation time on this `entering sequence' and show that for a given collection of particles  group sequence in some cases minimizes the transportation time better than a random sequence. We also introduce the `sorting cost' necessary to transform a random sequence into a group sequence and show that when this is included a random sequence can become advantageous in some conditions. We obtain these results not only from numerical simulations but also by theoretical analyses that generalize the simulation results for some special cases.
\end{abstract}

\maketitle


\section{INTRODUCTION}
Transportation systems are key topics in social or biological systems~\cite{appert2017transport}. In social systems, researchers have sought to obtain smooth transportation in various situations, such as production flow~\cite{croom2000supply,ezaki2015dynamics}, vehicular traffic~\cite{taniguchi2015jam, PhysRevE.87.062818,  PhysRevE.91.062818, yamamoto2017velocity}, and pedestrian evacuation~\cite{christensen2008agent, pelechano2008evacuation, yanagisawa2009introduction, ezaki2012simulation}. On the other hand, for biological systems, intracellular transportation along microtubules has been vigorously investigated~\cite{rothman1994mechanisms,muller2008tug, bressloff2013stochastic}.

Among various transportation models, the asymmetric simple exclusion process (ASEP), pioneered by MacDonald and Gibbs~\cite{macdonald1968kinetics, macdonald1969concerning}, has attracted much attention. It is a stochastic process on a one-dimensional lattice in which particles move asymmetrically. A derivative of ASEP, in which particles are allowed to hop unidirectionally (left to right in the present study) is called a totally asymmetric simple exclusion process (TASEP). In the field of nonequilibrium statistical mechanics, researchers have applied TASEP to various transportation problems, such as molecular-motor traffic~\cite{parmeggiani2003phase, chowdhury2005physics, chou2011non, appert2015intracellular}, vehicular traffic~\cite{RevModPhys.73.1067, PhysRevE.91.062818, PhysRevE.89.042813, tsuzuki2018effect,PhysRevE.87.062818, yamamoto2017velocity}, and the exclusive-queuing process~\cite{yanagisawa2010excluded, arita2010exclusive, arita2015exclusive}, especially since the TASEP with open boundary conditions has been solved exactly~\cite{PhysRevE.59.4899, 1751-8121-40-46-R01, 0305-4470-26-7-011}. 

In practice, researchers struggle to achieve smooth operation for various tasks, smooth logistics for various products, and an effective evacuation method for pedestrians in various situations, such as exit plans from sports stadiums and concert venues. To attain smooth flow in such situations, we often consider the sequence in which we perform tasks and pedestrians move because this sequence may affect the total performance of the systems. For example, slow pedestrians may block fast ones at the back of a narrow street, which worsens pedestrian flow. To investigate how the abovementioned sequences affect pedestrian flow in various systems, herein, we propose a modified TASEP comprising a finite number of multi-species particles, in which the entering sequences of the particles are considered.

In the proposed model, we consider the number of particles to be finite and study the transportation times of those particles. Note that we do not consider the steady state of the system itself. Minemura {\it et.al}~\cite{minemura2010productivity} investigated the transportation time for a hopping probability that depends upon the lot size, using the single-species TASEP with a finite number of particles. Other related works~\cite{1742-5468-2008-06-P06009, cook2009competition, 1742-5468-2009-02-P02012, 1742-5468-2012-05-P05008} also adopted a finite number of particles. In those models, however, particles circulated through a system comprising a lattice and a particle pool while the input or output rate was varied.

Additionally, the concept of multiple particles has already been extensively studied ~\cite{angel2006stationary,ferrari1994invariant,arita2006phase, mallick1996shocks, derrida1999bethe, ayyer2009two, privman2005nonequilibrium,ferrari2007stationary, cantini2008algebraic, uchiyama2008two, crampe2015open,aneva2003deformed,arita2009spectrum,duchi2005combinatorial,evans1997exact, krug1996phase,mallick1999exact,crampe2016integrable, arita2013matrix, prolhac2009matrix, ayyer2017exact, arndt1998spontaneous, evans1996bose, sasamoto2000one}. 
For example, second class particles were introduced in Refs.~\cite{angel2006stationary,ferrari1994invariant,arita2006phase, mallick1996shocks, derrida1999bethe, ayyer2009two, privman2005nonequilibrium,ferrari2007stationary, cantini2008algebraic, uchiyama2008two, crampe2015open} and more than two-species particles were introduced in Refs.~\cite{aneva2003deformed,arita2009spectrum, duchi2005combinatorial,evans1997exact, krug1996phase,mallick1999exact, crampe2016integrable, arita2013matrix, prolhac2009matrix,ayyer2017exact}. However, most of these studies focused on mathematically exact solutions to the systems under consideration by using such as Matrix Product Ansatz and did not much consider the application of the studied model to real-world situations. Furthermore, owing to their simplicity, periodic-boundary conditions have been adopted in many studies~\cite{angel2006stationary, arita2009spectrum, derrida1999bethe, arndt1998spontaneous, evans1996bose, evans1997exact, krug1996phase, sasamoto2000one, mallick1999exact, prolhac2009matrix, arita2013matrix}. Studies on multi-species ASEP with open boundaries and random updating were undertaken only recently~\cite{arita2006phase, ayyer2009two, privman2005nonequilibrium, uchiyama2008two, crampe2016integrable, crampe2015open, ayyer2017exact}; for example, Ref.~\cite{ayyer2017exact} obtained the exact phase diagram for a multi-species (more than two-species) ASEP. The present investigation primarily focuses on the problem of minimizing the transportation time, adopting open-boundary conditions and parallel updating. With the same boundary conditions and updating rules as the present study, Ref.~\cite{bengrine1999simulation} adopted particles with disorder, whereas jumping particles were introduced in Ref.~\cite{duchi2008combinatorial}. Note that majority related works considering multi-species particles assume that swapping between different types of particles, i.e., bidirectional particle hopping, can occur, whereas our model prohibits swapping
\footnote{
To be precise, if we regard vacant sites as particle 0, swapping between particle 0 and other kinds of particles (1, 2, 3,......) occurs. However, swapping between particles 1, 2, 3,...... does not occur.
}.

In contrast, to the best of our knowledge, no TASEP investigations that focus on the entering sequence of the particles (the key highlight of our model) have been reported thus far. Herein, we have considered two special types of sequences in particular: `random sequences' and `group sequences,' and we have compared the transportation times for these two types of sequences.  
In association with the entering sequence, we have introduced the sorting cost in our model. Without sorting, particles are typically transported at random, i.e., in a random sequence. Therefore, considering the cost of sorting particles from a random sequence into a group sequence is useful. In the present study, we define this sorting cost and compare the results obtained with and without sorting.

We have determined the dependence of the transportation time on the entering sequence of the particles from numerical simulations based on our model. Moreover, we find that the optimal sequence can vary, depending upon choice of parameter set, when the sorting cost is considered. In addition, we have succeeded in obtaining mathematical proofs of the simulation results for some special cases. 

The remainder of the present study is organized as follows. Section \ref{sec:model} describes the details of our proposed model and some important parameters, modifying the original TASEP. In Sec. \ref{sec:simulation}, we present and discuss the results of numerical simulations using the modified TASEP. Section \ref{sec:theory} presents theoretical analyses of the simulation results for some special cases. The paper concludes in Sec. \ref{sec:conclusion}.

\section{MODEL DESCRIPTION}
\label{sec:model}
\subsection{Original (single-species) TASEP with open-boundary conditions}
\begin{figure}[htbp]
\begin{center}
\includegraphics[width=8cm,clip]{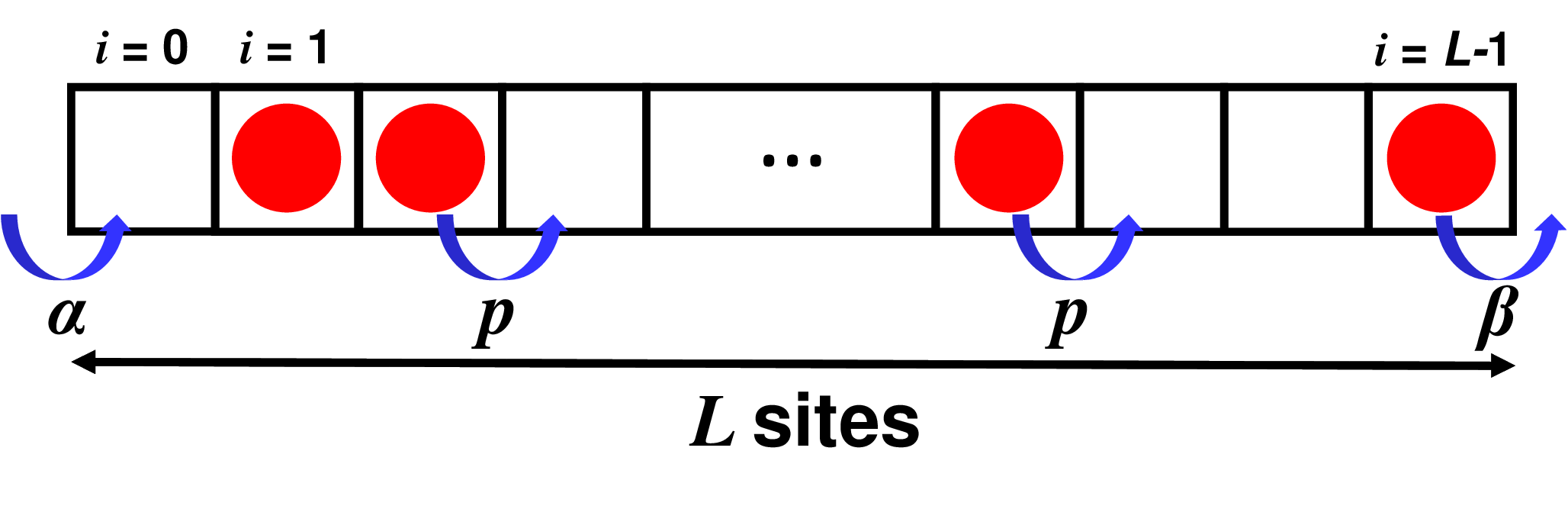}
\caption{(Color online) Schematic illustration of the original TASEP with open-boundary conditions.}
\label{fig:original_TASEP}
\end{center}
\end{figure}

The original TASEP with open-boundary conditions is defined as a one-dimensional lattice of $L$ sites, labeled from left to right $i=0,1,......,L-1$ (see Fig. \ref{fig:original_TASEP}). Each site can be either empty or occupied by a single particle. In the present study, we adopt discrete time steps and parallel updating. In parallel updating, the states of all the particles on the lattice are determined simultaneously in the next time step.
Notably, we can use random updating, which is usually adopted in the ASEP; however, we intentionally adopt parallel updating in the present study (see the specific reasons in \footnote{This system can be simulated also with random updating; however, we adopt parallel updating for the following three reasons. First, we consider that the results will be mainly applied to real systems, where particles and agents move simultaneously. It is generally natural to adopt parallel updating in the contexts of pedestrian dynamics or traffic flows~\cite{nagel1992cellular,burstedde2001simulation}.
Second, it is more natural to consider that time for swapping two particles is compatible with a one-time step with parallel updating than with random-sequential updating. Finally, the simulation times can be suppressed drastically with parallel updating compared to with random updating.}).
Particles enter the lattice from the left boundary with probability $\alpha$, and leave the lattice from the right boundary with probability $\beta$. In the bulk of the lattice, if the right-neighboring site is empty, a particle hops to that site with probability $p$; otherwise it remains at its present site. Our modified TASEP differs from this original one in the following four ways.

\subsection{Difference 1: Finite number of particles}
\begin{figure}[htbp]
\begin{center}
\includegraphics[width=8cm,clip]{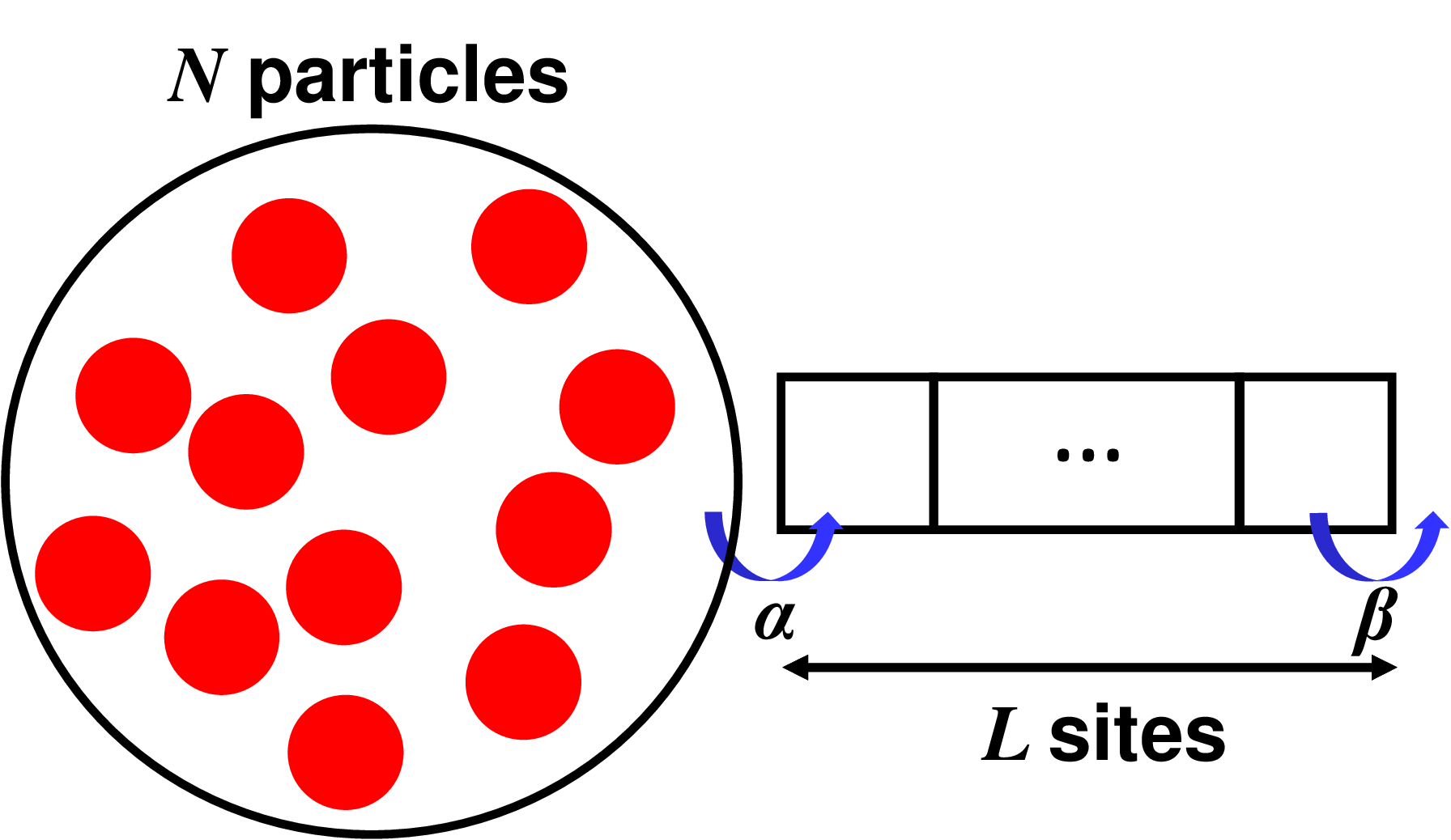}
\caption{(Color online) Schematic illustration of TASEP with a finite number of particles. This figure shows the case $N=12$.}
\label{fig:N}
\end{center}
\end{figure}

First, the number of particles $N$ is finite, as illustrated in Fig. \ref{fig:N}. The system evolves until the $N$th particle leaves the lattice. We define the transportation time $T$ as the time gap between the start of the simulation and the time when the $N$th particle leaves the lattice. 

\subsection{Difference 2: Multi-species particles}
\begin{figure}[htbp]
\begin{center}
\includegraphics[width=8cm,clip]{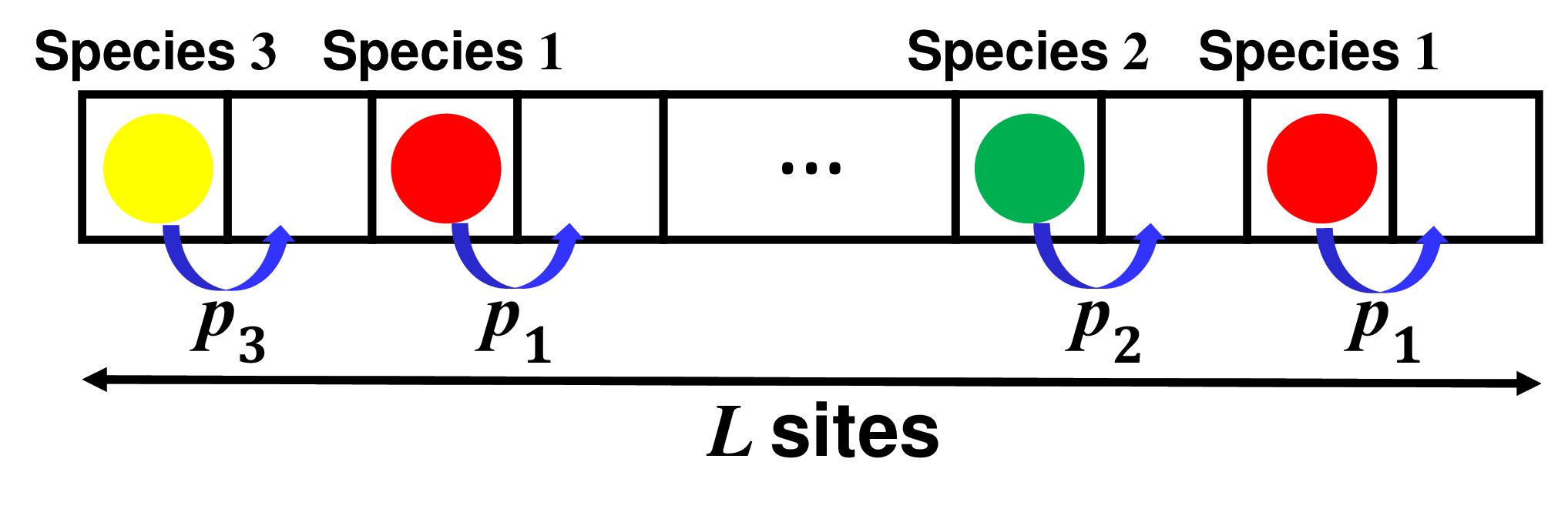}
\caption{(Color online) Schematic illustration of TASEP with multi-species particles. In this figure, we show a case with $S=3$, where the red particles belong to species 1, green ones to species 2, and the yellow ones to species 3.}
\label{fig:specie}
\end{center}
\end{figure}

Second, our model adopts multi-species particles, i.e., particles with different hopping probabilities, as illustrated in Fig. \ref{fig:specie}. Specifically, each of the $N$ particles is allocated to one of $S$ species, where $1\leq S\leq N$. Particles that belong to each species $s$ ($s=1,2,......,S$) all have the same hopping probability $p=p_s$ ($0< p_s\leq 1$). Note that with $S=1$ our model reduces to the single-species TASEP, whereas with $S=N$ all particles have different hopping probabilities. The fraction of all the $N$ particles allocated to each species $s$ is defined as $r_s$, obviously satisfying $\sum_{s=1}^S {r_s} = 1$.

\subsection{Difference 3: Consideration of entering sequence of particles}
\begin{figure}[htbp]
\begin{center}
\includegraphics[width=8cm,clip]{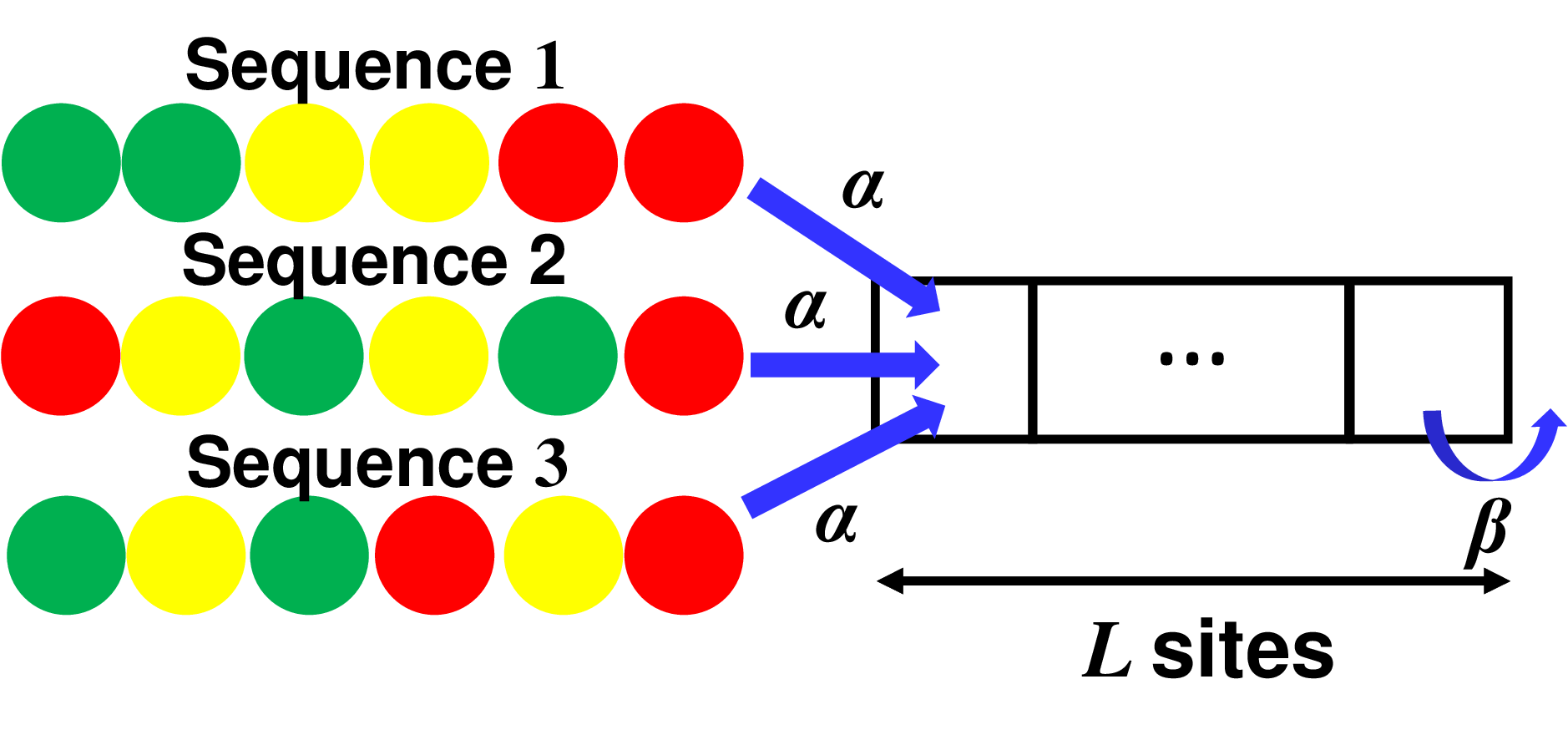}
\caption{(Color online) Schematic illustration of the entering sequences of particles. In this figure, we show three examples among all 90 \{= 6!/(2!2!2!)\} possible sequences for the case $N=6$, $S=3$, and $r_1=r_2=r_3=1/3$. Note that Sequence 1 is one example of a group sequence, whereas the others are examples of random sequences.}
\label{fig:sequence}
\end{center}
\end{figure}

\begin{figure}[htbp]
\begin{center}
\includegraphics[width=8cm,clip]{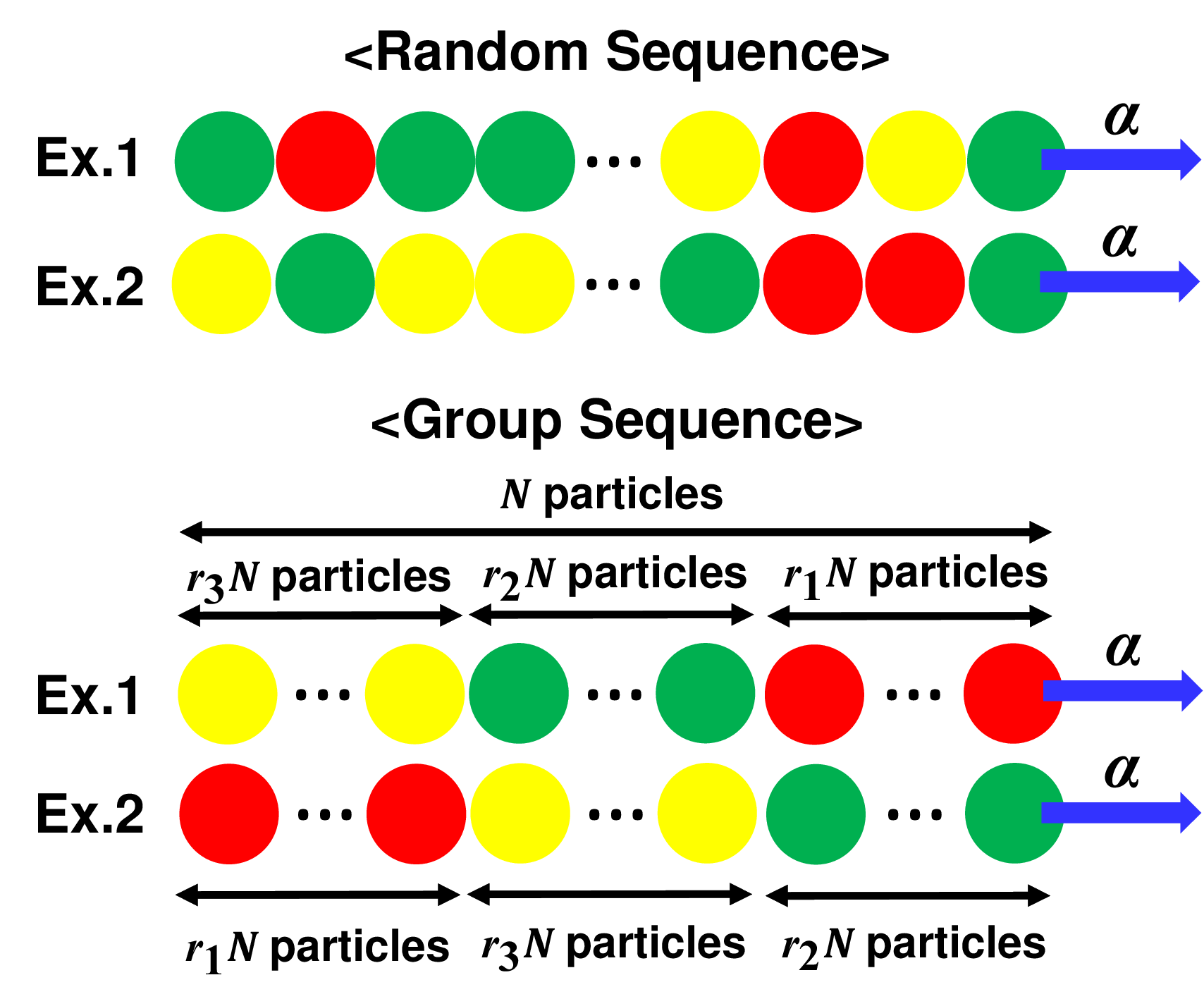}
\caption{(Color online) Schematic illustration of random (upper panel) and group (lower panel) sequences for $S=3$, where the red particles belong to species 1, the green ones to species 2, and the yellow ones to species 3. In the upper panel, we show two possible examples out of all $N!/\prod_{s=1}^3 (r_s N)!$ possible random sequences, whereas in the lower panel, we display two examples of all 6 (= 3!) possible group sequences. Note that in each case $r_1+r_2+r_3=1$.}
\label{fig:group_random}
\end{center}
\end{figure}

Third, we consider the sequence in which the particles enter the lattice (i.e., the `entering sequence'), which is the most important feature in our model. Specifically, particles form a queue before the left boundary and enter the lattice according to the sequence, as illustrated in Fig. \ref{fig:sequence}. In the present study, we investigate two types of sequences: `random sequences' and `group sequences,' as illustrated in Fig. \ref{fig:group_random}. 

In a random sequence, particles line up randomly regardless of their hopping probabilities. A random sequence thus has $N!/\prod_{s=1}^S (r_s N)!$ patterns. Note that in real situations without any controls, random sequences can be assumed to occur spontaneously.

On the other hand, in a group sequence, particles form groups of the same species and line up group by group. There are $S!$ possible patterns of group sequences, which are clearly among the random sequences. 

For the case $S=N$, where all hopping probabilities are different, we bunch the particles with similar hopping probabilities close together with each other as much as possible, imaginarily considering them as `continuous groups.' Consistent with this idea, we define a group sequence with $S=N$ as either an ascending or a descending sequence. Note that we define such a sequence by considering the rightmost particle to be the first particle in the sequence.

We define the transportation times for the random and group sequences to be $T_{\rm R}$ and $T_{\rm G}$, respectively. 

\begin{figure}[htbp]
\begin{center}
\includegraphics[width=9cm,clip]{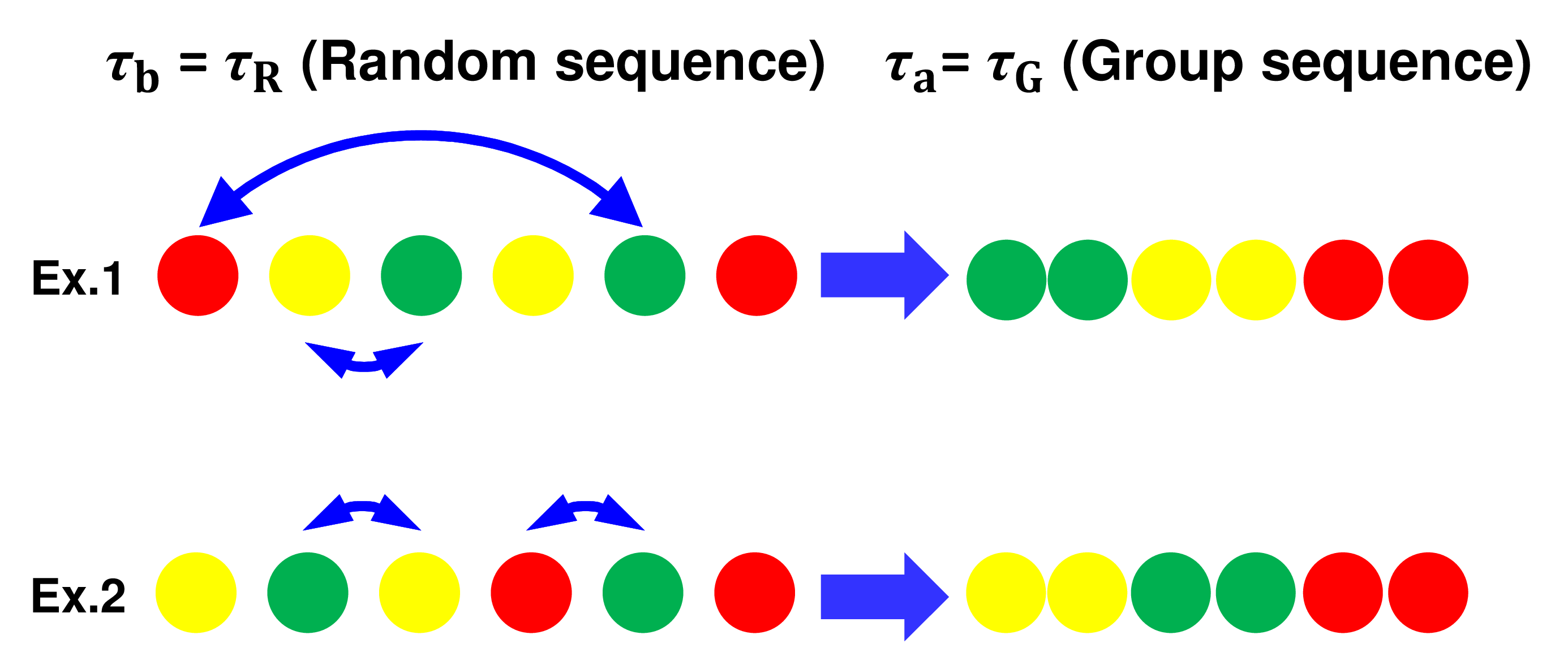}
\caption{(Color online) Two examples with $K(\tau_{\rm a}, \tau_{\rm b})=2$ for the case $N=6$ and $S=3$. For each sequence $\tau_{\rm R}$, we chose the one of all $6\ (=3!)$ possible sequences $\tau_{\rm G}$ so that $K(\tau_{\rm a}, \tau_{\rm b})$ is minimized.}
\label{fig:schematicsort}
\end{center}
\end{figure}

\subsection{Difference 4: Introduction of the sorting cost}
Finally, we introduce the cost of sorting the particles and investigate the effect of the sorting cost on the transportation time. Here, we define the sorting cost as the minimal number of exchanges $K(\tau_{\rm a}, \tau_{\rm b})$ necessary to sort the particles form sequence $\tau_{\rm b}$ to sequence $\tau_{\rm a}$, where $\tau_{\rm a}$ and $\tau_{\rm b}$ represent the sequence after sorting and before sorting, respectively. Note that the arguments of $K(\tau_{\rm a}, \tau_{\rm b})$ will be abbreviated in obvious cases. 

In the present study, $\tau_{\rm a}$ ($\tau_{\rm b}$) correspond to $\tau_{\rm G}$ ($\tau_{\rm R}$), where $\tau_{\rm R}$ and $\tau_{\rm G}$ represent a random sequence and a group sequence, respectively. The sequence $\tau_{\rm G}$ can differ depending upon $\tau_{\rm R}$; that is, $\tau_{\rm G}$ is determined so that the number of exchanges is minimized for each $\tau_{\rm R}$. Figure \ref{fig:schematicsort} shows two examples for which $K(\tau_{\rm a}, \tau_{\rm b})=2$ when $N=6$ and $S=3$. Note that we do not consider the distance between the exchanged particles.

We define the number of time steps necessary to sort the particles to be $\lambda K$, where the parameter $\lambda$ is the ratio of the sorting cost to number of TASEP time steps.

\section{SIMULATION RESULTS}
\label{sec:simulation}
In this section, we use numerical simulations to investigate the dependence of the transportation time on the entering sequence of the particles. 

\begin{figure*}[htbp]
\centering
\includegraphics[width=17.5cm,clip]{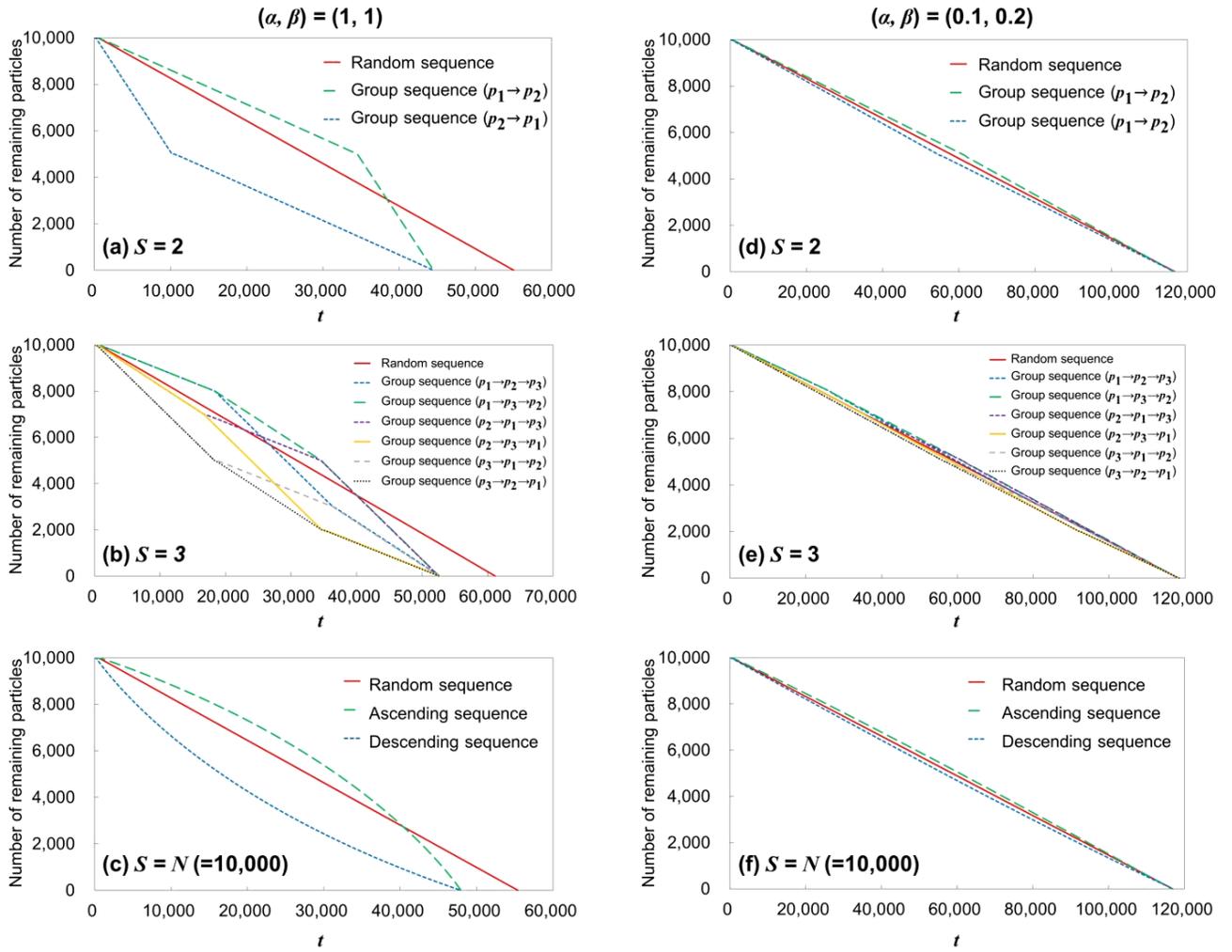}
\caption{(Color online) Simulation values of the number of particles remaining at time $t$ with $\lambda=0$ for (a) $S=2$, (b) $S=3$, and (c) $S=N$ with $(\alpha,\beta)=(1,1)$, and for (d) $S=2$, (e) $S=3$, and (f) $S=N$ with $(\alpha,\beta)=(0.1,0.2)$. For $S=2$, $S=3$, and $S=N$, respectively, we set $(p_1,p_2;r)=(0.5,1;0.5)$, $(p_1,p_2,p_3;r_1,r_2,r_3)=(0.4,0.6,0.8;0.2,0.3,0.5)$, and $p_s=1-0.6(N-s)/(N-1) \ (s=1,2,......, N)$, respectively, fixing $\lambda=0$. The notation '$p_s\to p_t$' means that a group of species $s$ is followed by a group of species $t$.}
\label{fig:transport}
\end{figure*}

\begin{figure*}[htbp]
\flushleft
\includegraphics[width=17.7cm,clip]{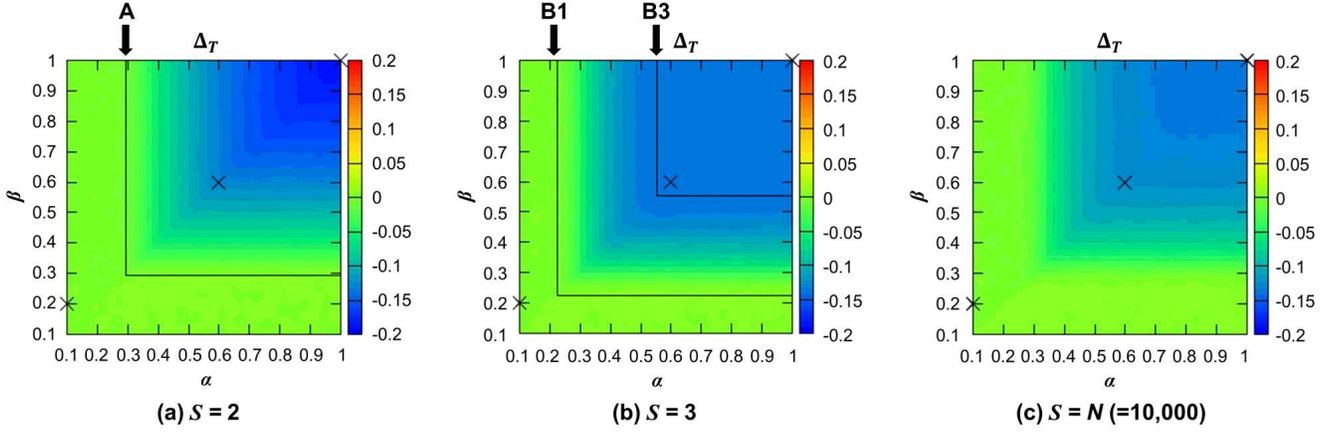}
\caption{(Color online) Simulation values of $\Delta_T$ for various $(\alpha,\beta)$ with (a) $S=2$, (b) $S=3$, and (c) $S=N$. The parameters other than $(\alpha,\beta)$ are the same as in Fig. \ref{fig:transport}. Note that three black crosses in each panel represent $(\alpha, \beta)=$(0.1, 0.2), (0.6, 0.6), and (1, 1), respectively. The color scale at the right of each panel represents the value of $\Delta_T$.}
\label{fig:deltaT}
\end{figure*}

\begin{figure*}[htbp]
\flushleft
\includegraphics[width=17.5cm,clip]{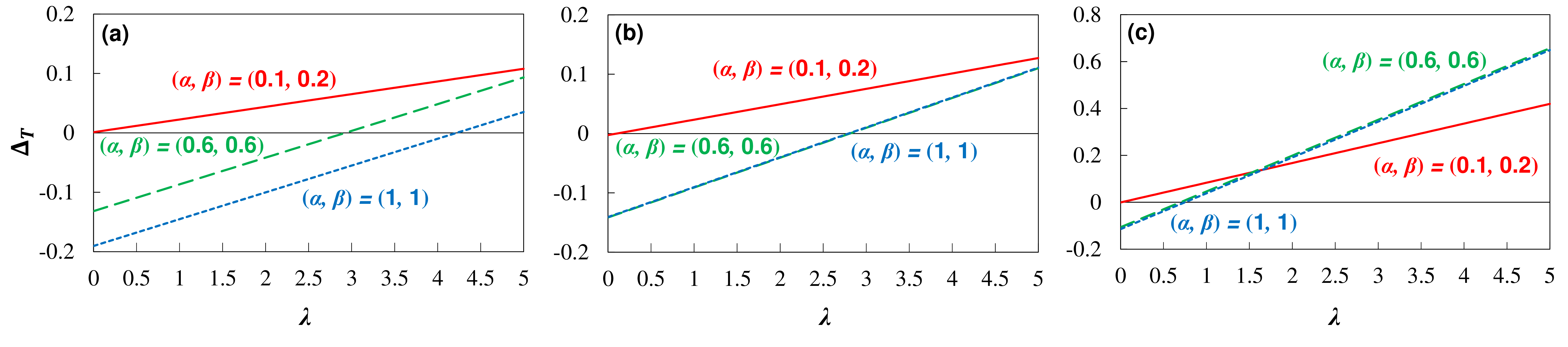}
\caption{(Color online) Simulation values of $\Delta_T$ with (a) $S=2$, (b) $S=3$, and (c) $S=N$ as functions of $\lambda$ for various $(\alpha,\beta) \in \{(0.1,0.2), (0.6,0.6), (1,1)\}$. The parameters other than $(\alpha,\beta)$ are the same as in Fig. \ref{fig:transport}.}
\label{fig:lambda}
\end{figure*}

\begin{figure*}[htbp]
\flushleft
\includegraphics[width=17.5cm,clip]{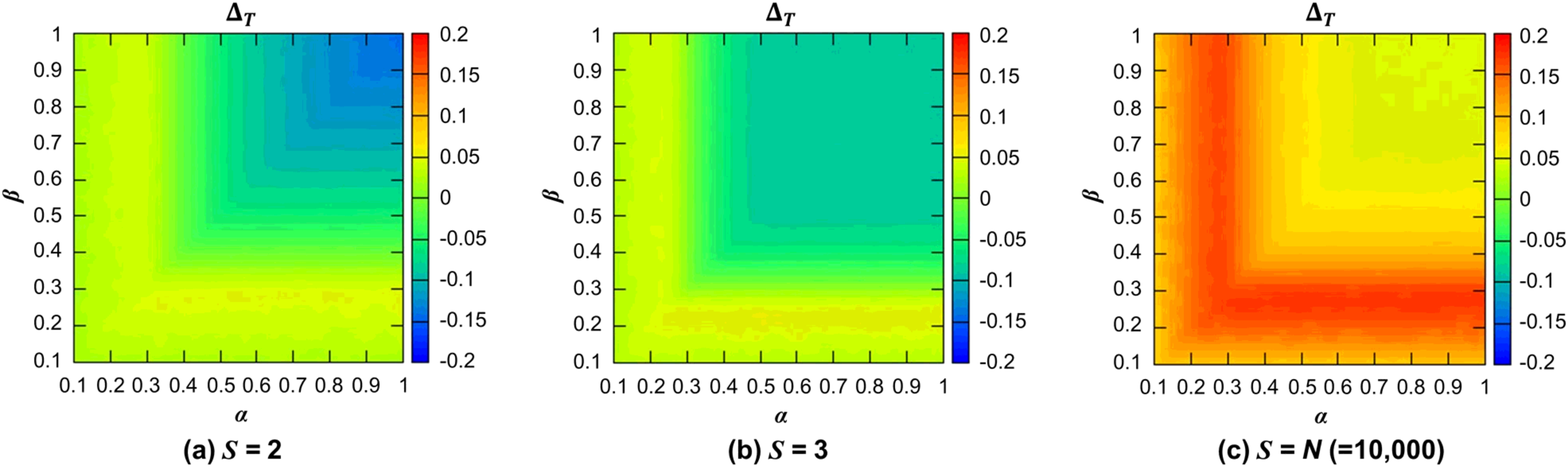}
\caption{(Color online) Simulation values of $\Delta_T$ for various $(\alpha,\beta)$ with (a) $S=2$, (b) $S=3$, and (c) $S=N$ for $\lambda=1$. The other parameters are the same as in Fig. \ref{fig:transport}.}
\label{fig:withsort}
\end{figure*}

In all the simulations below, we set $L=200$ and $N=10,000$; we validate this selection of $L$ an $N$ in Appendix \ref{sec:validity1}. We determine the value of $T$ for each parameter, and average $T$ over 100 trials for Fig. \ref{fig:transport} and over 10 trials for Figs. \ref{fig:deltaT}, \ref{fig:lambda}, and \ref{fig:withsort}). 

\subsection{Without sorting cost ($\lambda=0$)}
In this subsection, we set $\lambda=0$, i.e., we do not include the sorting cost. 

In Fig. \ref{fig:transport} we plot the simulation values of the number of particles that have not yet exited the lattice at time $t$ for $S\in\{2, 3, N\}$. We fix $(\alpha,\beta)=(1,1)$ for (a)--(c) and $(\alpha,\beta)=(0.1,0.2)$ for (d)--(f). In the figures, we refer to the number of particles that have not yet exited the lattice at time $t$ simply as the `remaining particles.' The simulation starts at $t=0$, and the number of particles becomes 0, i.e., the $N$th particle exits the lattice, at $t=T$. 

We note two important phenomena in Figs. \ref{fig:transport} (a)--(c). First, surprisingly, $T_{\rm G}$ is smaller than $T_{\rm R}$ for all three values of $S$ when $\alpha=\beta=1$. This result implies that the group sequences yield smoother transportation than the random ones for the cases $(\alpha,\beta)=(1,1)$.
Second, $T_{\rm G}$ seems not to depend upon the order of each group in the group sequence, which can take $S!$ possible patterns. 

On the other hand, in Figs. \ref{fig:transport} (d)--(f), unlike the cases in Figs. \ref{fig:transport} (a)--(c), the difference between $T_{\rm R}$ and $T_{\rm G}$ seems to vanish. 

In order to compare the difference between $T_{\rm R}$ and $T_{\rm G}$ for various $(\alpha,\beta)$, we define $\Delta_T$ as the ratio of the change from $T_{\rm R}$ to $T_{\rm G}$; that is,
\begin{equation}
\Delta_T=\frac{T_{\rm G}-T_{\rm R}}{T_{\rm R}}.
\end{equation}
From this definition of $\Delta_T$, $\Delta_T<0$ $(\Delta_T>0)$ indicates that group (random) sequences are preferable for smooth transportation. Note that in the following, to calculate $\Delta_T$ we assume that each group in a group sequence is arranged in ascending order in terms of species number $s$.

The simulation values of $\Delta_T$ for various $(\alpha,\beta)$ with (a) $S=2$, (b) $S=3$, and (c) $S=N$ are plotted in Fig. \ref{fig:deltaT}. Note that the black lines represent the boundaries between the low-density/high-density (LD/HD) and the maximal current (MC) phases of the single-species TASEP with hopping probability $p_1$ (boundary A) in Fig. \ref{fig:deltaT} (a), and $p_1$ (boundary B1) and $p_3$ (boundary B3) in Fig. \ref{fig:deltaT} (b), respectively.

Figure \ref{fig:deltaT} shows that for all three values of $S$, $\Delta_T$ is small in the region where ${\rm min}(\alpha,\beta)$ is relatively large. [In Fig. \ref{fig:deltaT} (b), $\Delta_T$ finally yields to a constant value in the upper-right region beyond boundary B3.] On the other hand, $\Delta_T$ is small in the region where ${\rm min}(\alpha,\beta)$ is relatively small. [In Fig. \ref{fig:deltaT} (a) and (b), $\Delta_T$ is almost 0, especially in the lower-left region beyond the boundary A or B1.]  Here, we term the region with $\Delta_T<0$ as the `group-advantageous region' ($T_{\rm R}>T_{\rm G}$), whereas we designate the region with $\Delta_T\approx0$ as a `neutral region' ($T_{\rm R}\approx T_{\rm G}$), if it exists.

These results indicate that group sequences can make transportation smoother than random sequences when the system is mainly governed by the bulk region of the lattice, but the dependence on the type of sequences vanishes (or decreases) when the system is mainly governed by the boundaries.

\subsection{With sorting cost ($\lambda>0$)}
In this subsection, we consider the sorting cost by varying $\lambda$ for the same parameter sets in the previous subsection. Appendix \ref{sec:sortalgo} presents specific schemes for obtaining the minimal number of exchanges necessary to sort the particles in the simulations.

Figure \ref{fig:lambda} plots $\Delta_T$ for (a) $S=2$, (b) $S=3$, and (c) $S=N$ as functions of $\lambda$ for various $(\alpha,\beta) \in \{(0.1, 0.2), (0.6, 0.6), (1, 1)\}$, which are plotted as black crosses in Fig. \ref{fig:deltaT}. We emphasize again that in the region with $\Delta_T<0$ group sequences are preferred, even if when the sorting cost is considered, whereas in the region with $\Delta_T>0$ random sequences are preferred. Note that the cases with $\lambda=0$ correspond to those obtained without considering the sorting cost. 

As discussed in the previous subsection, we note that $\Delta_T\leq 0$ for almost all $(\alpha,\beta)$ when $\lambda=0$, indicating that sorting is almost always beneficial for smooth transportation. However, once the sorting cost is considered, the sign of $\Delta_T$ can become positive, especially in the region where ${\rm min}(\alpha,\beta)$ is relatively small, indicating that sorting is not always beneficial. Note that the curves of $(\alpha,\beta)=$(0.6, 0.6) and (1, 1) are observed to overlap each other in Fig. \ref{fig:lambda} (b) and (c), unlike Fig. \ref{fig:lambda} (a). This happens because there is no difference in $\Delta_T$ at these two points when $\lambda = 0$, as we can see in Figs. \ref{fig:deltaT} (b) and (c).

Figure \ref{fig:withsort} plots $\Delta_T$ for (a) $S=2$, (b) $S=3$, and (c) $S=N$ for various $(\alpha,\beta)$ with $\lambda=1$. In this figure, we note the existence of a new region in which $\Delta_T> 0$, which we term a `random-advantageous region' ($T_{\rm R}<T_{\rm G}$). This new region widens as $\lambda$ increases, finally resulting in the complete disappearance of the group-advantageous region for large enough $\lambda$. Note that Fig. \ref{fig:withsort} (c) exhibits only a random-advantageous region.

\section{THEORETICAL ANALYSES}
\label{sec:theory}
In this section, we show that the simulation results can be theoretically reproduced in some special cases. Specifically, we have succeeded in obtaining a mathematical proof of the appearance of the group-advantageous region for any group number $S(>1)$ when $\lambda=0$.

\subsection{Approximate flow of a multi-species TASEP}
In this subsection, before calculating $T$, we briefly discuss the steady-state flow $Q_S$ of the multi-species TASEP that corresponds to a random sequence. We write $Q_S=Q_S(p_1,......,p_S; r_1,......,r_S)$, with the arguments abbreviated in obvious cases. When the flow $Q$ is simulated for each parameter set, we first evolve the system for $10^5$ time steps and then average over the next $10^6$ time steps.

\begin{figure*}[htbp] \centering
$
\begin{pmatrix}
P_{00}\\
P_{0\ast}\\
P_{10}\\
P_{1\ast}\\
P_{20}\\
P_{2\ast}
\end{pmatrix}
=
\begin{pmatrix}
1-\alpha & (1-\alpha)\beta & 0 & 0 & 0 & 0 \\
0 & (1-\alpha)(1-\beta) & p_1 & 0 & p_2 & 0 \\
r\alpha & r\alpha\beta & 1-p_1 & \beta & 0 &0 \\
0 & r\alpha(1-\beta) & 0 & 1-\beta & 0 & 0\\
(1-r)\alpha & (1-r)\alpha\beta & 0 & 0 & 1-p_2 & \beta \\
0 & (1-r)\alpha(1-\beta) & 0 & 0 & 0 & 1-\beta \\
\end{pmatrix}
\begin{pmatrix}
P_{00}\\
P_{0\ast}\\
P_{10}\\
P_{1\ast}\\
P_{20}\\
P_{2\ast}
\end{pmatrix}
\ \ \ (2) 
$
\end{figure*}

\begin{figure*}[htbp]
\flushleft
\includegraphics[width=17.5cm,clip]{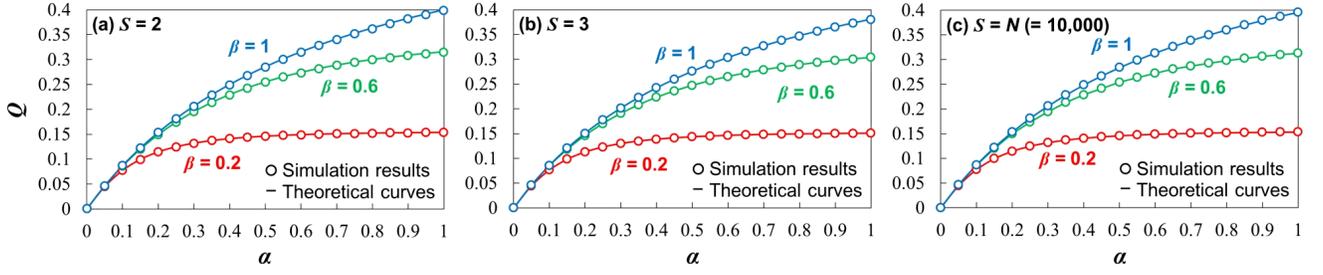}
\caption{(Color online) Simulation (circles) and theoretical (curves) values of (a) $Q_2$, (b) $Q_3$, and (c) $Q_N$ for $L=2$ as functions of $\alpha$ for various $\beta \in \{0.2, 0.6, 1\}$. The other parameters are fixed at (a) $(p_1,p_2;r)=(0.5,1;0.5)$, (b) $(p_1,p_2,p_3;r_1,r_2,r_3)=(0.4,0.6,0.8;0.2,0.3,0.5)$, (c) $p_s=1-0.6(S-s)/(S-1)$ ($s=1,2,......, N$) and $r_s=1/N(=1/10,000)$.}
\label{fig:L2}
\end{figure*}

\subsubsection{$L=2$}
\label{sec:L2}

This subsection presents the derivation of an approximate $Q_S$ based on a Markov chain model. Due to the difficulty of considering general values of $L$ (the length of the lattice) and $S$ (the number of particle species), we consider the simplest case---with $L=2$ and $S=2$. As two species of particles exist---that is, particles with hopping probability $p_1$ and particles with $p_2$---each site may have three states: `unoccupied (state 0),' `occupied by a particle 1 (state 1),' and 'occupied by a particle 2 (state 2).' This results in 9 possible states; however, noting that it is not necessary to distinguish the particle at site 1 because it always leaves the lattice with probability $\beta$, the number of possible states can be reduced to 6. Here, we define the probability distribution $P_{ij}$ ($i=0,1,2, j=0,\ast$), where $i$ and $j$ represent the state number of site 0 and 1, respectively. Note that state $\ast$ indicates either of state 1 or 2.


The master equations for the steady state are summarized in Eq. (2), using the relation $r_1(= r) + r_2 = 1$. Note that $r_1$ and $r_2$ are replaced with $r$ and $1-r$, respectively, for the case $S=2$. In addition, $P_{ij}$ must satisfy the normalization condition
\setcounter{equation}{2}
\begin{equation}
\sum_{i=0}^2  P_{i0} + \sum_{i=0}^2  P_{i\ast}= 1.
\label{eq:normal}
\end{equation}

From Eqs. (2) and (\ref{eq:normal}), the flow of the system can be written as a function of $p_1$, $p_2$, and $r$; that is, $Q_2(p_1, p_2; r)$, is given by the following expression:

\begin{equation}
\begin{split}
Q_2(p_1, p_2; r)&=p_1P_{10}+p_2P_{20}\\
&=\frac{p_1p_2 A}{\{(1-r)p_1+r p_2\}A+p_1p_2B},
\end{split}
\end{equation}
where

\begin{equation}
A=\alpha\beta(\alpha+\beta-\alpha\beta)
\end{equation}
and
\begin{equation}
B=\alpha^2+\beta^2-\alpha^2\beta-\alpha\beta^2+\alpha\beta.
\end{equation}
The specific forms of the probability distributions are summarized in Appendix \ref{sec:probdis}.

For $r=1$ and $p_1=p$, the system reduces to the single-species TASEP with the flow $Q_1(p)$, where
\begin{equation}
Q_1(p)=\frac{p A}{pB+A}.
\end{equation}
Note that the flow of the single-species TASEP for general $L$ is exactly solved in Ref.~\cite{evans1999exact}.
Therefore, assuming that the value $p=p_{\rm h}$ satisfies the condition $Q_2(p_1, p_2; r)=Q_1(p)$, we can derive
\begin{equation}
p_{\rm h}=\frac{p_1 p_2}{(1-r)p_1+r p_2}.
\label{eq:ph}
\end{equation} 
The quantity $p_{\rm h}$ is termed the harmonic mean of $p_1$ and $p_2$. Accordingly, for $L=2$, $Q_2$ is equivalent to $Q_1(p=p_{\rm h})$. This relation holds for any species number $S(>2)$, as we show in Appendix \ref{sec:generals}. 

Figure \ref{fig:L2} compares the simulation and theoretical curves for various $\beta \in \{0.2, 0.6, 1\}$ with (a) $S=2$, (b) $S=3$ and (c) $S=N(=10,000)$. In all the figures, the simulations show very good agreement with our exact analyses.

\begin{figure}[htbp]
\begin{center}
\includegraphics[width=8cm,clip]{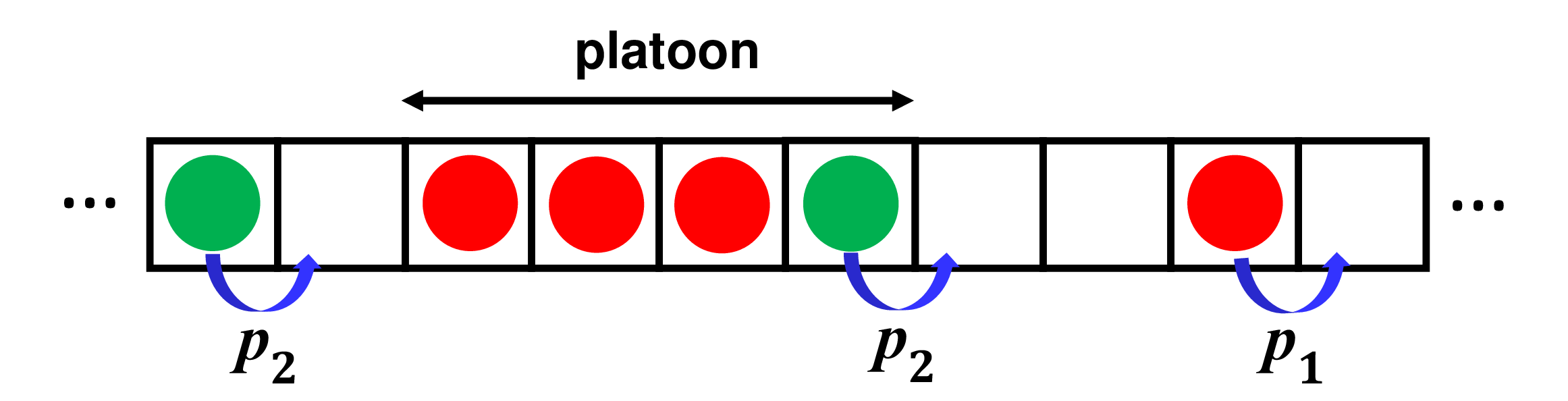}
\caption{(Color online) Schematic illustration of a platoon. In this figure, we set $S=2$, with the red particles belonging to species 1 (faster) and the green ones to species 2 (slower). A green particle blocks the red particles behind it, so that the trailing red particles cannot hop with probability $p_1$ but only with probability $p_2$, which is less than $p_{\rm h}$.}
\label{fig:platoon}
\end{center}
\end{figure}

\begin{figure*}[htbp]
\flushleft
\includegraphics[width=17.5cm,clip]{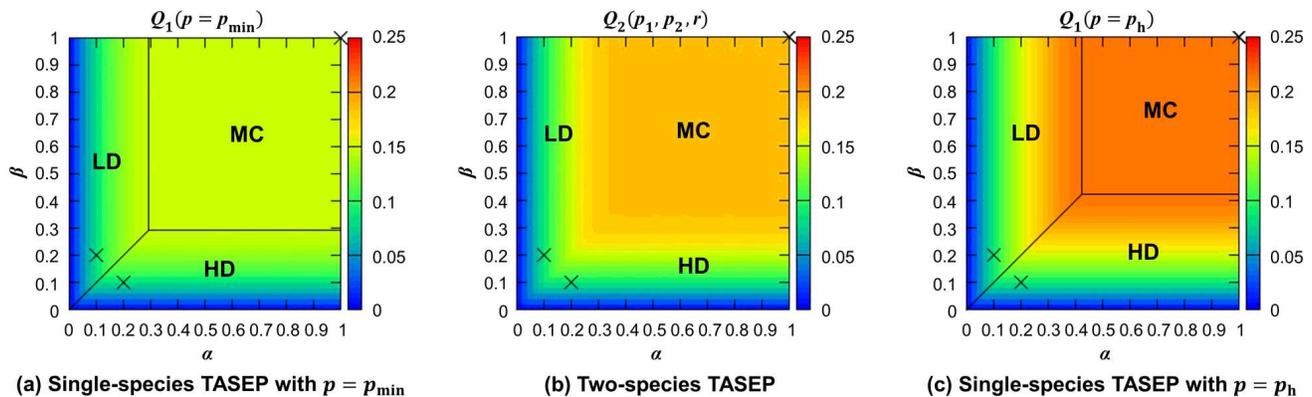}
\caption{(Color online) Phase diagrams. The color bars indicate the simulation values of (a) $Q_1(p=p_{\rm min})$, (b) $Q_2(p_1,p_2;r)$, and (c) $Q_1(p=p_{\rm h})$, respectively, obtained by fixing $(p_1,p_2;r)=(0.5,1;0.5)$, $p_{\rm min}=0.5$, and $p_{\rm h}=1/(0.5/0.5+0.5/1)=2/3$. Note that the three black crosses in each figure represent $(\alpha, \beta)=$(0.1, 0.2), (0.2, 0.1), and (1,1), respectively.}
\label{fig:phasediagram}
\end{figure*}

\begin{figure}[htbp]
\begin{center}
\includegraphics[width=6.5cm,clip]{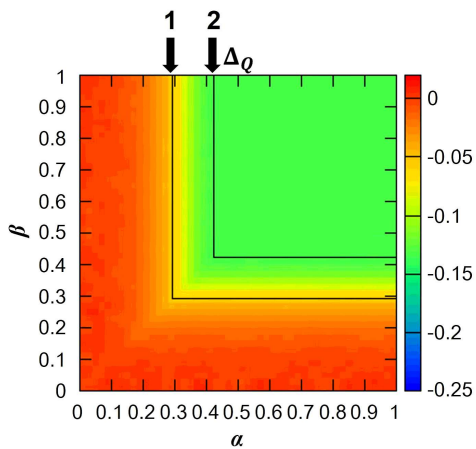}
\caption{(Color online) Calculated values of $\Delta_Q$ for various $(\alpha, \beta)$, fixing $(p_1,p_2;r)=(0.5, 1; 0.5)$.}
\label{fig:deltaQphase}
\end{center}
\end{figure}

\begin{figure*}[htbp]
\flushleft
\includegraphics[width=17.5cm,clip]{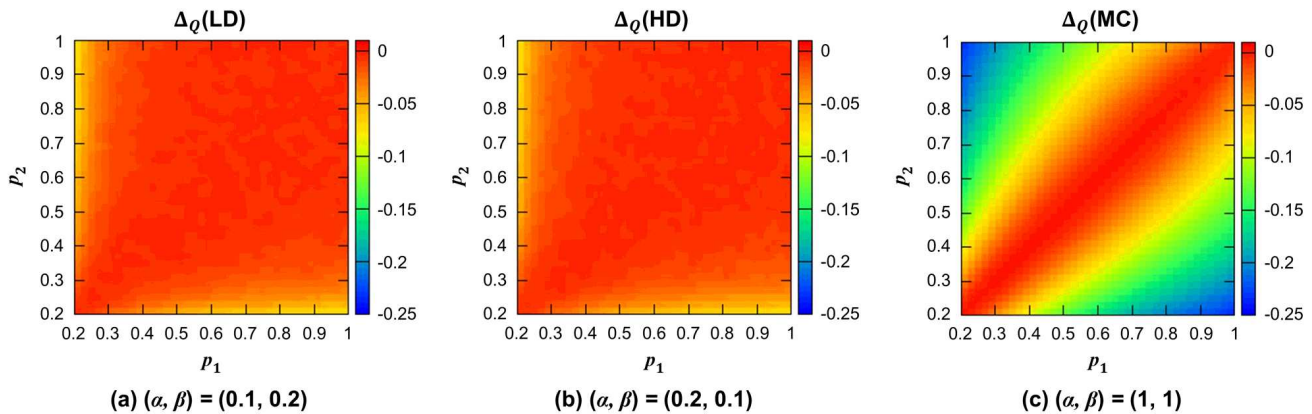}
\caption{(Color online) Calculated values of $\Delta_Q$ for various $(p_1, p_2)$, $S=2$, and $r=0.5$, (a) $(\alpha, \beta)=(0.1, 0.2)$ (LD), (b) $(\alpha, \beta)=(0.2, 0.1)$ (HD), and (c) $(\alpha, \beta)=(1, 1)$ (MC).}
\label{fig:Q2}
\end{figure*}

\begin{figure*}[htbp]
\flushleft
\includegraphics[width=17.5cm,clip]{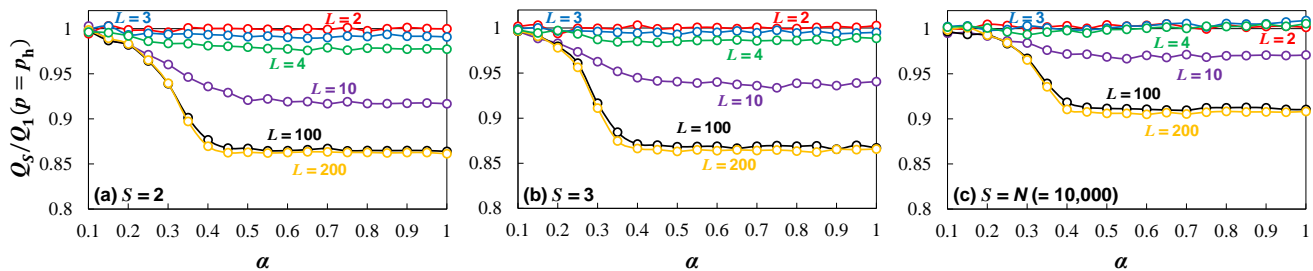}
\caption{(Color online) Simulation values of the ratio (a) $Q_2/Q_1(p=p_{\rm h})$, (b) $Q_3/Q_1(p=p_{\rm h})$, and (c) $Q_N/Q_1(p=p_{\rm h})$ as functions of $\alpha$ for various $L \in \{2, 3, 4, 10, 100, 200\}$, fixing $\beta=0.6$. The other parameters are fixed at (a) $(p_1,p_2;r)=(0.5,1;0.5)$, (b) $(p_1,p_2,p_3;r_1,r_2,r_3)=(0.4,0.6,0.8;0.2,0.3,0.5)$, (c) $p_s=1-0.6(S-s)/(S-1)$ ($s=1,2,......, N$) and $r_s=1/N(=1/10,000)$.}
\label{fig:QLdepend}
\end{figure*}

\subsubsection{General $L(>2)$}
\label{sec:generall}
For general $L$ and $S$, it is complicated to solve the master equations. Therefore, in this subsection, we instead assume an inequality, based on the results in the previous subsection and the qualitative discussions, and confirm the validity of the inequality by the simulations.

First, for general $L$ and $S$, $Q_S$ is clearly larger than $Q_1(p=p_{\rm min})$, where $p_{\rm min}={\rm min}\{p_1,p_2,......,p_S\}$. 

In addition, for $L>2$, a platoon can be observed in the bulk of the lattice, in which a slower particle behaves as a bottleneck, and faster particles behind it cannot hop with a probability larger than that of the smaller one, i.e., less than $p_{\rm h}$, as shown in Fig. \ref{fig:platoon}. This phenomenon suppresses the flow, implying that $Q_S$ is smaller than $Q_1(p=p_{\rm h})$.

Consequently, $Q_S$ satisfies the following inequality;
\begin{equation}
Q_1(p=p_{\rm min})< Q_S < Q_1(p=p_{\rm h}).
\label{eq:QLQSQH}
\end{equation}
In this subsection, we hereafter consider the case $S=2$. 

Figure \ref{fig:phasediagram} shows the phase diagrams obtained by plotting the simulation values for (a) the single-species TASEP with $p=p_{\rm min}$, (b) the two-species TASEP, and (c) the single-species TASEP with $p=p_{\rm h}$, respectively. Note that $Q_2$ ($Q_1$) are the simulation (theoretical) values (and similarly hereafter). 

Comparing these three figures shows that Eq. (\ref{eq:QLQSQH}) obviously holds. In addition, as in Figs. \ref{fig:phasediagram} (a) and (c), we find that three different phases---HD, LD, and MC---also exist in Fig. \ref{fig:phasediagram} (b). Due to Eq. (\ref{eq:QLQSQH}), the boundaries between the LD (HD) and MC phases of Fig. \ref{fig:phasediagram} (b) lie between those of Figs. \ref{fig:phasediagram} (a) and \ref{fig:phasediagram} (c). Note that the black lines in Figs. \ref{fig:phasediagram} (a) and (c) are theoretical boundaries, based on the fact that the boundary between the LD and MC phases of the single-species TASEP with hopping probability $p$~\cite{PhysRevE.59.4899} can be written as
\begin{eqnarray}
\left\{\begin{array}{ll}
\alpha=1-\sqrt{1-p} \ \land \ 1-\sqrt{1-p}<\beta<1 \\
\beta=1-\sqrt{1-p} \ \land \ 1-\sqrt{1-p}<\alpha<1
\end{array} \right..
\end{eqnarray}

Here, as for $\Delta_T$, we define $\Delta_Q$ as the ratio of the change from $Q_1(p=p_{\rm h})$ to $Q_2$; that is,
\begin{equation}
\Delta_Q=\frac{Q_2-Q_1(p=p_{\rm h})}{Q_1(p=p_{\rm h})},
\end{equation}
and we note that $\Delta_Q=0$ when $Q_1(p=p_{\rm h})=0$.

Figure \ref{fig:deltaQphase} shows $\Delta_Q$ for various ($\alpha, \beta$), for the fixed parameter set $(p_1,p_2;r)=(0.5, 1; 0.5)$. The black lines represent the boundaries between the LD/HD and MC phases of the single-species TASEP with hopping probability $p_1$ (boundary 1) and the single-species TASEP with $p=p_{\rm h}$ (boundary 2). Therefore, the lower-left (upper-right) region beyond boundary 1 (boundary 2) corresponds to the LD/HD (MC) phases both for the two-species TASEP and for the single-species TASEP with $p=p_{\rm h}$. This figure confirms that $\Delta_Q$ starts from 0 in the LD/HD phase, decreases, and finally yields to a constant value in the MC phase as $(\alpha,\beta)$ approaches the upper right.

Figure \ref{fig:Q2} plots $\Delta_Q$ as a function ($p_1, p_2$) for various $(\alpha,\beta)\in\{(0.1, 0.2), (0.2, 0.1), (1, 1)\}$, fixing $r=0.5$. 
Note that both the single-species TASEP with hopping probability $p_{\rm h}$ and the two-species TASEP exhibit the LD, HD, and MC phases with $(\alpha,\beta)=$(0.1, 0.2), (0.2, 0.1), and (1, 1), respectively. This is because $\forall (p_1,p_2)$ $(0.2\leq p_1,p_2\leq 1)$ \ $0.1< 1-\sqrt{1-p_{\rm min}}\leq 1-\sqrt{1-p_{\rm h}}$. For example, Fig. \ref{fig:phasediagram} confirms that those three points exist within each corresponding phase for $(p_1,p_2)=(0.5, 1)$. 

In Figs. \ref{fig:Q2} (a) and (b), we find that $\Delta_Q$ is approximately 0, whereas $\Delta_Q$ deviates from 0 in Fig. \ref{fig:Q2} (c), as is also observed in Fig. \ref{fig:deltaQphase}. These phenomena can be explained as follows.

First, in the LD/HD phase, $Q_2$ is mainly governed by the input/output probability, leading to $\Delta_Q\to 0$, i.e., $Q_2$ approaches $Q_1(p=p_{\rm h})$. This is because $Q_2$ deviates from $Q_1(p=p_{\rm h})$ mainly due to the existence of platoons, which do not influence the flow much in this phase. Note that $\Delta_Q$ decreases as $\alpha$ or $\beta$ approaches 0.2, because the influence of platoons increases, approaching the MC phase of the two-species TASEP.

On the other hand, in the MC phase, $Q_2$ is mainly governed by the bulk region of the lattice. Therefore, the existence of platoons has a more critical influence on $Q_2$, causing $\Delta_Q$ to deviate from 0; i.e., $Q_2<Q_1(p=p_{\rm h})$. Especially as $|p_1-p_2|$ increases, the extent of the deviation also increases. This is because the effect of platoons increases when there is a large gap between $p_1$ and $p_2$.

Figure \ref{fig:QLdepend} plots (a) $Q_2/Q_1(p=p_{\rm h})$, (b) $Q_3/Q_1(p=p_{\rm h})$, and (c) $Q_N/Q_1(p=p_{\rm h})$ for various $L \in \{2, 3, 4, 10, 100, 200\}$. Both of $Q_S$ ($S=2,3,10,000$) and $Q_1(p=p_{\rm h})$ are obtained by the simulations. In all the figures, we observe that for $L>2$, $Q_S/Q_1(p=p_{\rm h})$ generally becomes less than 1, i.e., $Q_S<Q_1(p=p_{\rm h})$, especially when $\alpha$ increases, i.e., the system approaches and exhibits the MC phase. Note that for $S=10,000$, the difference among $L=2, 3, 4$ is unclear; however, it becomes obvious when $L\geq 10$. Those results imply that Eq. (9) and its qualitative discussions can be applicable for general $L$ and $S$.

\subsection{Relation between $T_{\rm R}$ and $T_{\rm G}$ \\ without the sorting cost}
\label{sec:TRTG}
Hereafter, we assume $p_1<p_2<......<p_S$, $S>1$, $\forall s \ r_s>0$, and $\alpha>0$.

In this subsection, we fix $\lambda=0$, i.e., we do not consider the sorting cost. If for any number of particle species $s$, $r_{s}N$ is large enough for $T_{\rm G}$ and $T_{\rm R}$ to be determined by the steady-state flow (see Appendix \ref{sec:validity2}), we obtain

\begin{equation}
T_{\rm R}\approx \frac{N}{Q_S}
\label{eq:Tr}
\end{equation}
and
\begin{equation}
T_{\rm G}\approx \sum_{s=1}^S \frac{r_{s}N}{Q_1(p=p_s)}.
\label{eq:Tg}
\end{equation}
Note that this approximation immediately implies the independence of $T_{\rm G}$ from the order of the group sequence. Strictly speaking, $T_{\rm G}$ can differ depending on the order of each group in the group sequence. However, that difference can be ignored for large $N$ (see Appendix \ref{sec:validity2}).  

In addition, we define the transportation times of the particles with the same hopping probabilities $p_{\rm h}$ and $p_{\rm min}$ as

\begin{equation}
T_{\rm H}\approx \frac{N}{Q_1(p=p_{\rm h})}
\label{eq:Th}
\end{equation}
and
\begin{equation}
T_{\rm M}\approx \frac{N}{Q_1(p=p_{\rm min})},
\label{eq:Tm}
\end{equation}
respectively. From Eqs. (\ref{eq:QLQSQH}), (\ref{eq:Tr}), (\ref{eq:Th}) and (\ref{eq:Tm}), we immediately obtain the inequality
\begin{equation}
T_{\rm H}< T_{\rm R} <T_{\rm M}.
\label{eq:TMTRTH}
\end{equation}

In the following, we show that a general relation between $T_{\rm R}$ and $T_{\rm G}$ can be obtained mathematically for general $S(>1)$. We emphasize that this relation can be proven by comparing $T_{\rm H}$ and $T_{\rm G}$, and not by comparing $T_{\rm R}$ and $T_{\rm G}$ directly. 
Here, we introduce the new function $f(\alpha,\beta; p_1,......,p_S; r_1,......, r_S)$, which is defined as follows:

\begin{equation}
f(\alpha,\beta; p_1,......,p_S; r_1,......, r_S)=T_{\rm H}-T_{\rm G}.
\label{eq:f}
\end{equation}
Because we can assume $\alpha<\beta$ without loss of generality, we adopt this assumption in the following discussion, writing in abbreviated form $f(\alpha,\beta; p_1,......,p_S; r_1,......, r_S)=f(\alpha)$. Note that for cases with $\alpha\geq\beta$, the theoretical results can be obtained simply by replacing $\alpha$ (LD) with $\beta$ (HD).

A contour map of $f(\alpha)$ in the $(\alpha,\beta)$ plane exhibits four large regions, which are summarized in Tab. \ref{tab:region}.

\renewcommand{\arraystretch}{1.2}
\begin{table}[htbp]
\centering
\caption{Classification of Regions}
\label{tab:region}
\begin{tabular}{c|c}
Region No. &  Range  \\ \hline \hline
1 & $\alpha<1-\sqrt{1-p_1}$ \\ 
2 & $1-\sqrt{1-p_1}\leq\alpha<1-\sqrt{1-p_{\rm h}}$ \\ 
3 & $1-\sqrt{1-p_{\rm h}}\leq\alpha<1-\sqrt{1-p_S}$ \\ 
4 & $1-\sqrt{1-p_S}\leq\alpha$ \\ \hline
\end{tabular}
\end{table}
\renewcommand{\arraystretch}{1}
In the following subsections, we examine the behavior of $f(\alpha)$ according to this classification. 

\subsubsection{Region 1: $\alpha<1-\sqrt{1-p_1}$}
\label{sec:region1}
In this region, all the steady-state phases of the single-species TASEP for any $p_s$ exhibit the LD phase. Here, the steady-state flow for the single-species TASEP with parallel updating~\cite{PhysRevE.59.4899} is given by
\begin{eqnarray}
Q_1(p)=\left\{ \begin{array}{ll}
\displaystyle\alpha \frac{p-\alpha}{p-\alpha^2} & {\rm for \ LD \ phase.}  \\
\displaystyle\frac{1}{2} (1-\sqrt{1-p}) & {\rm for \ MC \ phase.}
\end{array} \right.
\label{eq:Q1}
\end{eqnarray}
Therefore, we obtain

\begin{equation}
T_{\rm G}\approx \sum_{s=1}^{S} \frac{r_s N(p_s-\alpha^2)}{\alpha(p_s-\alpha)}
\label{eq:Tgld}
\end{equation}
and
\begin{equation}
T_{\rm H} \approx \frac{N(p_{\rm h}-\alpha^2)}{\alpha(p_{\rm h}-\alpha)}.
\label{eq:Thld}
\end{equation}
From Eqs.(\ref{eq:Tgld}) and (\ref{eq:Thld}), we obtain $f(\alpha)$ as
\begin{equation}
f(\alpha)=\frac{N(p_{\rm h}-\alpha^2)}{\alpha(p_{\rm h}-\alpha)}-\sum_{s=1}^{S} \frac{r_s N(p_s-\alpha^2)}{\alpha(p_s-\alpha)}.
\label{eq:fregion1}
\end{equation}
After some calculations, we obtain
\begin{equation}
f(\alpha)<0;
\label{eq:fLD}
\end{equation}
the detailed derivation is given in Appendix \ref{sec:region1app}.

\subsubsection{Region 2: $1-\sqrt{1-p_1}\leq\alpha<1-\sqrt{1-p_{\rm h}}$}
\label{sec:region2}
This region is further divided into ($u-1$) subregions, as summarized in Tab. \ref{tab:region2}. 
\renewcommand{\arraystretch}{1.2}
\begin{table}[htbp]
\centering
\caption{Classification of subregions in Region 2}
\label{tab:region2}
\begin{tabular}{c|c}
Subregion No. &  Range  \\ \hline \hline
2--$1$ & $1-\sqrt{1-p_1}\leq\alpha<1-\sqrt{1-p_2}$ \\
2--$2$ & $1-\sqrt{1-p_u}\leq\alpha<1-\sqrt{1-p_{u+1}}$ \\ 
... & ...\\
2--$v$ & $1-\sqrt{1-p_{v}}\leq\alpha<1-\sqrt{1-p_{v+1}}$ \\
... & ...\\
2--$(u-1)$ & $1-\sqrt{1-p_{u-1}}\leq\alpha<1-\sqrt{1-p_{\rm h}}$ \\ \hline
\end{tabular}
\end{table}
\renewcommand{\arraystretch}{1}

In Subregion 2--$v$ ($v=1,2,......,u-1$), the single-species TASEP with $p=p_1,......, p_{v}, p_{\rm h}$ exhibits the MC phase, whereas that with $p=p_{v+1},......, p_{S}$ displays the LD phase.
Therefore, using Eq. (\ref{eq:Q1}), we obtain
\begin{eqnarray}
Q_1(p)=\left\{ \begin{array}{ll}
\displaystyle\frac{1}{2} (1-\sqrt{1-p}) & {\rm for} \ p=p_1,......,p_{v-1}, p_{\rm h}. \\
\displaystyle\alpha \frac{p-\alpha}{p-\alpha^2} & {\rm for} \ p=p_{v+1},......, p_{S}.
\end{array} \right.
\label{eq:Q1region2}
\end{eqnarray}
From Eqs. (\ref{eq:Tg}), (\ref{eq:Th}), and (\ref{eq:Q1region2}), $T_{\rm G}$ and $T_{\rm H}$ can be written as follows;

\begin{equation}
T_{\rm G}\approx \sum_{s=1}^{v} \frac{2r_s N}{1-\sqrt{1-p_s}}+\sum_{s=v+1}^{S} \frac{r_s N(p_s-\alpha^2)}{\alpha(p_s-\alpha)}
\label{eq:Tgregion2}
\end{equation}
and
\begin{equation}
T_{\rm H} \approx \frac{N(p_{\rm h}-\alpha^2)}{\alpha(p_{\rm h}-\alpha)}.
\label{eq:Thregion2}
\end{equation}
From Eqs. (\ref{eq:Tgregion2}) and (\ref{eq:Thregion2}), we thus obtain $f(\alpha)$ in the form
\begin{equation}
\begin{split}
f(\alpha)&\approx \frac{N(p_{\rm h}-\alpha^2)}{\alpha(p_{\rm h}-\alpha)}\\
&\quad-\sum_{s=1}^{v} \frac{2r_s N}{1-\sqrt{1-p_s}}-\sum_{s=v+1}^{S} \frac{r_s N(p_s-\alpha^2)}{\alpha(p_s-\alpha)}.
\end{split}
\label{eq:fregion2}
\end{equation}
For $1-\sqrt{1-p_{v}}\leq\alpha<1-\sqrt{1-p_{v+1}}$, due to $\alpha<p_{v+1}<......<p_S$ and $\alpha<p_{\rm h}$, we obtain $p_s-\alpha> 0$ for $s=v+1,......,S$ and $p_{\rm h}-\alpha> 0$. The function $f(\alpha)$ is continuous and differentiable with respect to $\alpha$, including at each boundary (see Appendix \ref{sec:differentiable}). However, the signs of $f(\alpha)$ and $df(\alpha)/d\alpha$ are not specified. Note that the following condition
\begin{equation}
\lim_{\alpha \to q_h-0} \frac{df(\alpha)}{d\alpha}>0,
\end{equation}
where $q_h=1-\sqrt{1-p_{\rm h}}$ indicates that $f(\alpha)$ increases monotonically at least near the boundary between Subregion 2--($u-1$) and Region 3. This is discussed in Appendix \ref{sec:region2app}.

\subsubsection{Region 3: $1-\sqrt{1-p_{\rm h}}\leq\alpha<1-\sqrt{1-p_S}$}
\label{sec:region3}
Similarly to Region 2, this region is further divided into ($S-u+1$) subregions, as summarized in Tab. \ref{tab:region3}. Note that Subregion 3--1 vanishes in the case $p_{\rm h}=p_u$, resulting in ($S-u$) subregions.

\renewcommand{\arraystretch}{1.2}
\begin{table}[htbp]
\centering
\caption{Classification of subregions in Region 3}
\label{tab:region3}
\begin{tabular}{c|c}
Subregion No. &  Range  \\ \hline \hline
3--$u$ & $1-\sqrt{1-p_{\rm h}}\leq\alpha<1-\sqrt{1-p_u}$ \\
3--$(u+1)$ & $1-\sqrt{1-p_u}\leq\alpha<1-\sqrt{1-p_{u+1}}$ \\ 
... & ...\\
3--$v$ & $1-\sqrt{1-p_{v-1}}\leq\alpha<1-\sqrt{1-p_{v}}$ \\
... & ...\\
3--$S$ & $1-\sqrt{1-p_{S-1}}\leq\alpha<1-\sqrt{1-p_{S}}$ \\ \hline
\end{tabular}
\end{table}
\renewcommand{\arraystretch}{1}

In Region 3--$v$ ($v=u,u+1,......,S$), the single-species TASEP with $p=p_1,......, p_{v-1}, p_{\rm h}$ exhibits the MC phase, whereas that with $p=p_{v},......, p_{S}$ displays the LD phase.
Therefore, using Eq. (\ref{eq:Q1}), we obtain $Q_1(p)$
\begin{eqnarray}
Q_1(p)=\left\{ \begin{array}{ll}
\displaystyle\frac{1}{2} (1-\sqrt{1-p}) & {\rm for} \ p=p_1,......,p_{v-1}, p_{\rm h}. \\
\displaystyle\alpha \frac{p-\alpha}{p-\alpha^2} & {\rm for} \ p=p_{v},......, p_{S}.
\end{array} \right.
\label{eq:Q1region3}
\end{eqnarray}
From Eqs. (\ref{eq:Tg}), (\ref{eq:Th}), and (\ref{eq:Q1region3}), $T_{\rm G}$ and $T_{\rm H}$ can be written as follows;

\begin{equation}
T_{\rm G}\approx \sum_{s=1}^{v-1} \frac{2r_s N}{1-\sqrt{1-p_s}}+\sum_{s=v}^{S} \frac{r_s N(p_s-\alpha^2)}{\alpha(p_s-\alpha)}
\label{eq:Tgregion3}
\end{equation}
and
\begin{equation}
T_{\rm H} \approx \frac{2N}{1-\sqrt{1-p_{\rm h}}}.
\label{eq:Thregion3}
\end{equation}
From Eqs. (\ref{eq:Tgregion3}) and (\ref{eq:Thregion3}), $f(\alpha)$ becomes
\begin{equation}
\begin{split}
f(\alpha)&\approx \frac{2N}{1-\sqrt{1-p_{\rm h}}}\\
&\quad-\sum_{s=1}^{v-1} \frac{2r_s N}{1-\sqrt{1-p_s}}-\sum_{s=v}^{S} \frac{r_s N(p_s-\alpha^2)}{\alpha(p_s-\alpha)}.
\end{split}
\label{eq:fregion3}
\end{equation}
For $1-\sqrt{1-p_{v-1}}\leq\alpha<1-\sqrt{1-p_{v}}$, due to $\alpha<p_{v}<......<p_S$, we obtain $p_s-\alpha>0$ for $s=v,......,S$.  Similarly to Region 2, $f(\alpha)$ is continuous and differentiable with respect to $\alpha$ including at each boundary. 

After some calculations,we find that $d f(\alpha)/d\alpha$ satisfies
\begin{equation}
\frac{d f(\alpha)}{d\alpha}>0
\label{eq:fprimeregion3}
\end{equation}
in each subregion, as discussed in detail in Appendix \ref{sec:region3app}. Eq. (\ref{eq:fprimeregion3}) indicates that $f(\alpha)$ is a monotonically increasing function of $\alpha$ throughout Region 3.  

\subsubsection{Region 4: $1-\sqrt{1-p_S}\leq\alpha$}
\label{sec:region4}
In this region, all the steady-state phases of the single-species TASEP for any hopping probabilities $p_s$ and $p_{\rm h}$ exhibit the MC region. Note that this region vanishes in the case $p_S=1$ because $1-\sqrt{1-p_S}=1$. 

Therefore, we obtain

\begin{equation}
T_{\rm G}\approx \sum_{s=1}^{S} \frac{2 r_s N}{1-\sqrt{1-p_s}}
\label{eq:Tg2}
\end{equation}
and
\begin{equation}
T_{\rm H}\approx \frac{2N}{1-\sqrt{1-p_{\rm h}}},
\label{eq:Th2}
\end{equation}
both of which are independent of $\alpha$.
From Eqs. (\ref{eq:Tg2}) and (\ref{eq:Th2}),we thus obtain $f(\alpha)$ as

\begin{equation}
f(\alpha)=\frac{2N}{1-\sqrt{1-p_{\rm h}}}-\sum_{s=1}^{S} \frac{2 r_s N}{1-\sqrt{1-p_s}}.
\label{eq:fregion4}
\end{equation}

After some calculations, we obtain
\begin{equation}
f(\alpha)>0,
\label{eq:fMC}
\end{equation}
the detailed derivation of which is given in Appendix \ref{sec:region4app}.

\subsubsection{Relation between $T_{\rm R}$ and $T_{\rm G}$}
With the results of Subsec. \ref{sec:region1}--\ref{sec:region4}, we can obtain a general relation between $T_{\rm R}$ and $T_{\rm G}$ for some special cases. Table \ref{tab:sign} summarizes the signs of $f(\alpha)$ and $df(\alpha)/d\alpha$ in each region. Note that `U' indicates that the sign is unclear.

\renewcommand{\arraystretch}{1.2}
\begin{table}[htbp]
\centering
\caption{Sign of $f(\alpha)$ and $df(\alpha)/d\alpha$}
\label{tab:sign}
\begin{tabular}{c|c|c}
Region No. & $f(\alpha)$ & $df(\alpha)/d\alpha$ \\ \hline \hline
1 & $-$ & U\\
2 & U & U\\ 
3 & U & $+$ \\
4 & $+$ & 0\\ \hline
\end{tabular}
\end{table}
\renewcommand{\arraystretch}{1}

Considering Tab. \ref{tab:sign} and the continuity of $f(\alpha)$ including at each boundary (see Appendix \ref{sec:differentiable}), we find from the intermediate value theorem that $\exists \alpha_{\rm cr}$ such that $f$ satisfies 
\begin{equation}
f(\alpha=\alpha_{\rm cr})=0\Leftrightarrow T_{\rm H}=T_{\rm G}
\end{equation}
in Region 2 or 3. The specific conditions that $\alpha_{\rm cr}$ must satisfy are given in Appendix \ref{sec:f0}.

Defining $\alpha_{\rm cr,max}$ as the largest value among the quantities $\alpha_{\rm cr}$, we obtain $f(\alpha)>0$---i.e., $T_{\rm  H}>T_{\rm G}$---in the region where $\alpha>\alpha_{\rm cr,m}$. This is because $f(\alpha)$ is continuous and increases monotonically from a point in Region 2 (and through Region 3), to yield $f(\alpha)>0$ in Region 4. 

Considering Eq. (\ref{eq:TMTRTH}), we finally obtain 
\begin{equation}
T_{\rm G}< T_{\rm H}< T_{\rm R}
\label{eq:TrTg}
\end{equation}
in the region $\alpha>\alpha_{\rm cr,m}$. Eq. (\ref{eq:TrTg}) means that $\Delta_T<0$, reproducing the simulation results in the region where ${\rm min}(\alpha,\beta)$ is relatively large. This result indicates that the group-advantageous region must appear even in a case with $p_S=1$, for which Region 4 vanishes.  

In analogy with the discussion above, we can also predict that a region with $\Delta_T<0$ must appear in the case $S=N$.

\subsection{Relation between $T_{\rm R}$ and $T_{\rm G}$ with sorting cost}
In this subsection, we discuss the change in the relation between $T_{\rm R}$ and $T_{\rm G}$ when $\lambda>0$, i.e., when the sorting cost is included. In the following, we first obtain a general formula for the sorting cost and then evaluate upper and lower limits to $\lambda$.
 
\subsubsection{General formula for the sorting cost}
\label{sec:K}
First, we calculate mathematically the averaged minimal number of exchanges necessary to sorting the particles from random to group sequences. 

We here define $\overline{K}$ as the averaged value of $K$, using the fact that $\tau_{\rm R}$ can take $N!/\prod_{s=1}^S (r_s N)!$ patterns with equal probability. 
We thus have
\begin{equation}
\overline{K}=\frac{\prod_{s=1}^S (r_s N)!}{N!}\sum_{\forall{\tau_{\rm R}}}K(\tau_{\rm G},\tau_{\rm R}).
\label{eq:overK}
\end{equation}

If $K^{\prime}(\tau_{\rm G},\tau_{\rm R})$ is the minimal number of exchanges necessary to sort the particles from a random sequence $\tau_{\rm R}$ to a given fixed group sequence $\tau_{\rm G}$, then $K^{\prime}(\tau_{\rm G},\tau_{\rm R})$ satisfies 
\begin{equation}
K^{\prime}(\tau_{\rm G},\tau_{\rm R})={\rm min}\{K(\tau_{\rm G},\tau_{\rm R})\}.
\label{eq:Kprime}
\end{equation}
Note that the number of elements of $\{K(\tau_{\rm G},\tau_{\rm R})\}$ is equal to that of $\{\tau_{\rm G}\}$ from the definition. Eqs. (\ref{eq:overK}) and (\ref{eq:Kprime}) indicate that the best group sequence $\tau_{\rm G}$ can vary depending on the particular random sequence $\tau_{\rm R}$. 

Due to the difficulty of a general calculation of $\overline{K}$, we instead calculate $\overline{K}^{\prime}$, which is defined as follows:
\begin{equation}
\overline{K}^{\prime}=\frac{\prod_{s=1}^S (r_s N)!}{N!}\sum_{\forall{\tau_{\rm R}}}K^{\prime}(\tau_{\rm G}, \tau_{\rm R}),
\end{equation}
where $\tau_{\rm G}$ is a fixed sequence out of the set \{$\tau_{\rm G}$\} for all possible $\tau_{\rm R}$.

For $S=2$ and $S=N$, $\overline{K}^\prime$ can be generally calculated as 
\begin{eqnarray}
\overline{K}^{\prime}=\left\{ \begin{array}{ll}
r(1-r)N & {\rm for} \ S=2, \\
N-\displaystyle\sum_{k=1}^N \frac{1}{k} & {\rm for} \ S=N,
\end{array} \right.
\end{eqnarray}
the detailed derivations of which are discussed in Appendix \ref{sec:sort}.

Figure \ref{fig:Kcompare} shows the ratio $\overline{K}/\overline{K}^\prime$ for (a) $S=2$ and (b) $S=N$. Both figures show that $\overline{K}/\overline{K}^\prime\approx1$, i.e., $\overline{K}\approx\overline{K}^\prime$, indicating that there is no problem in substituting $\overline{K}^\prime$ for $\overline{K}$ for large enough $N$. 

In the following calculations, we therefore use $\overline{K}^\prime$ instead of $\overline{K}$ because $\overline{K}^\prime$ can be represented by a general formula, whereas $\overline{K}$ cannot. 

\begin{figure}[htbp]
\begin{center}
\includegraphics[width=8cm,clip]{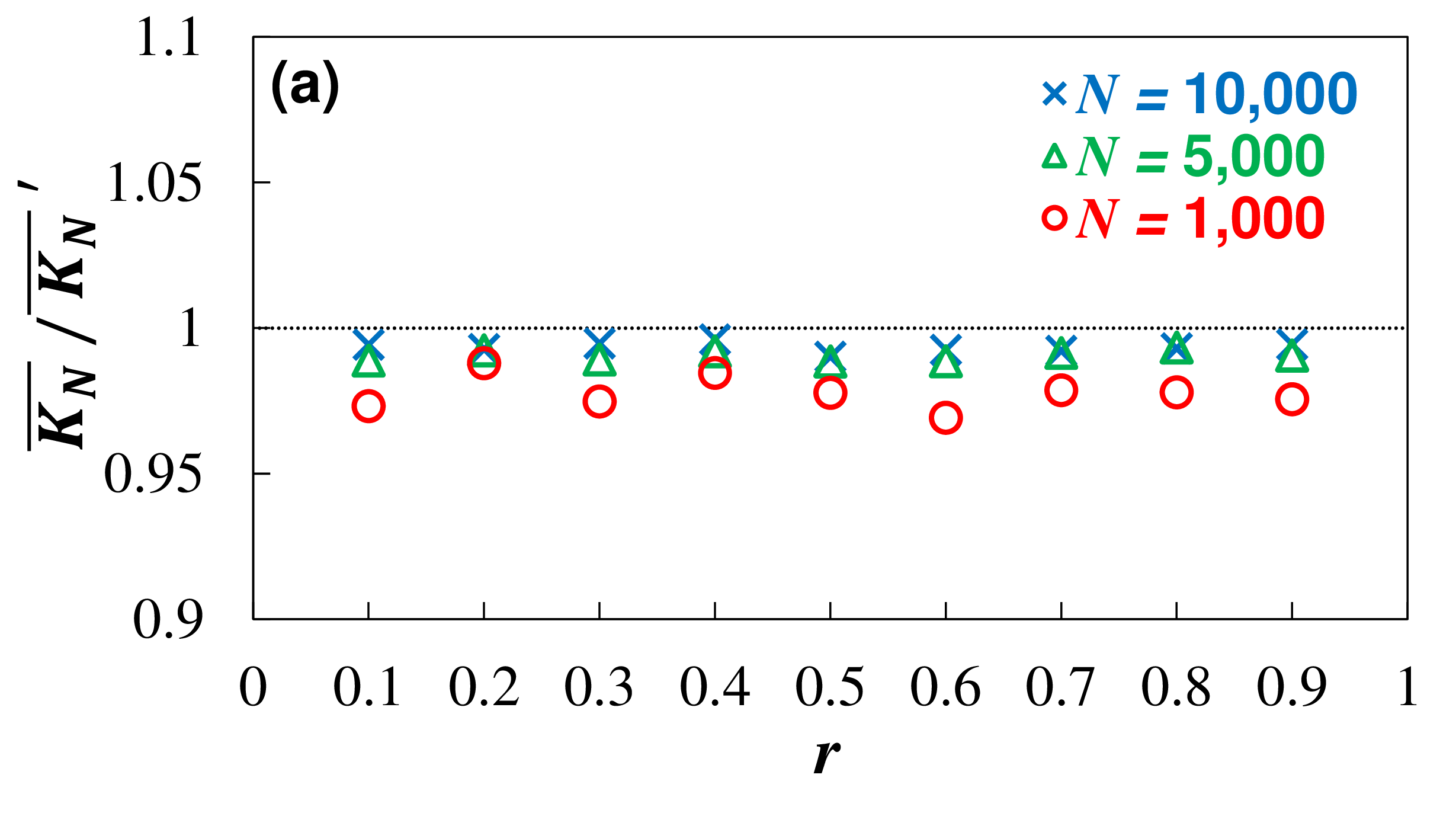}
\\
\includegraphics[width=8cm,clip]{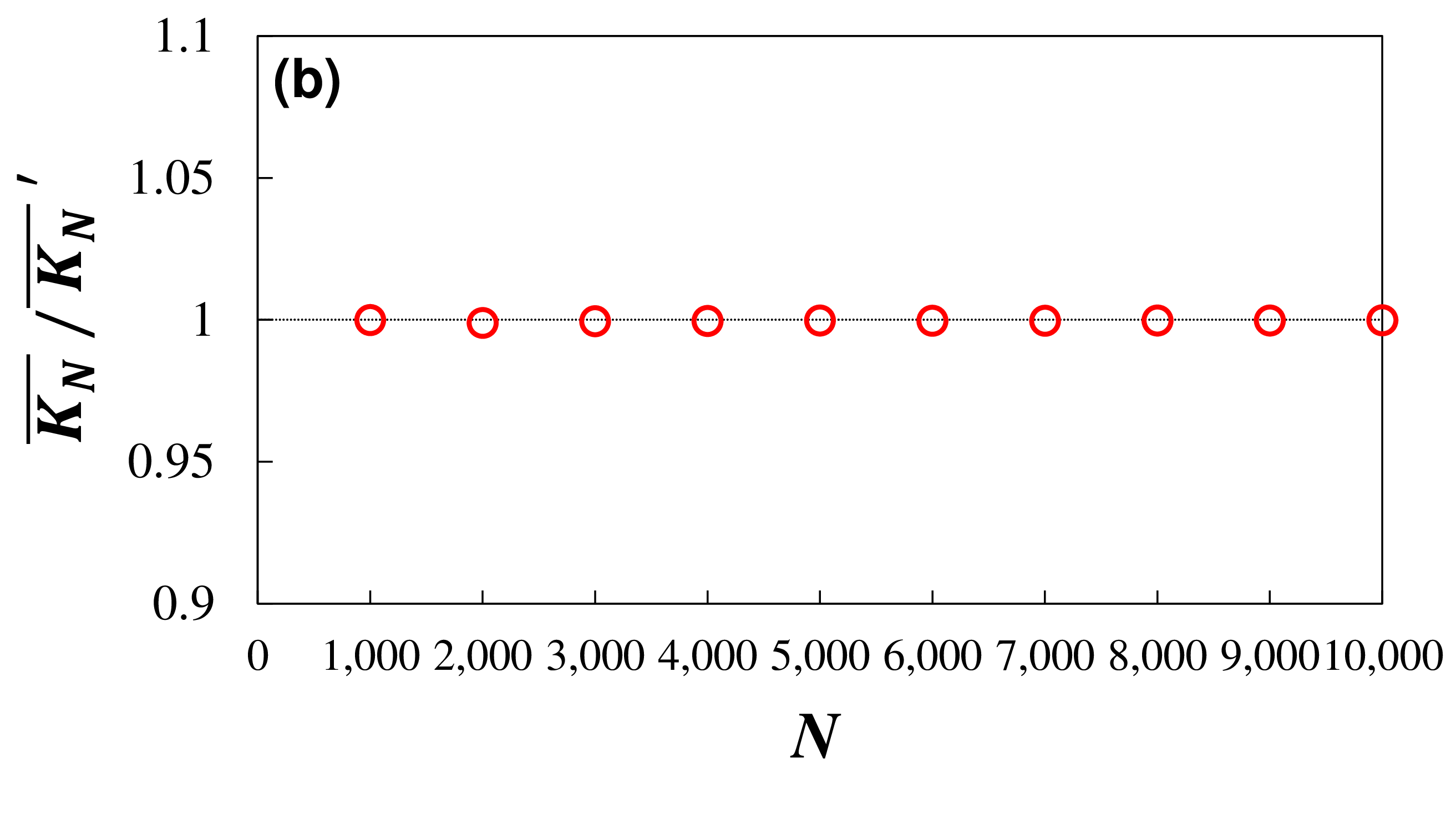}
\caption{(Color online) (a) Simulation values of the ratio $\overline{K}/\overline{K}^\prime$ for various $N\in\{1,000({\rm red}), 5,000({\rm green}), 10,000({\rm blue})\}$ for $S=2$. (b) Simulation values of the ratio $\overline{K}/\overline{K}^\prime$ for $S=N$. Note that $\overline{K}$ and $\overline{K}^\prime$ are each simulation values obtained by respectively averaging over 100 trials.}
\label{fig:Kcompare}
\end{center}
\end{figure}

\subsubsection{Upper and lower limits to $\lambda_{\rm cr}$}

We first define $\lambda=\lambda_{\rm cr}\geq 0$, as the value for which $T_{\rm R}=T_{\rm G}$. Note that $\lambda_{\rm cr}$ is defined to be equal to 0 if $T_{\rm R}\leq T_{\rm G}$ when $\lambda=0$. From the definition of $\lambda_{\rm cr}$, the random-advantageous region appears when $\lambda>\lambda_{\rm cr}$. Based on the discussions in Subsec. \ref{sec:TRTG} and \ref{sec:K}, we here evaluate $\lambda_{\rm cr}$ for $S=2$ and $\alpha<\beta$.  

To take into account the sorting cost, we add the term $\lambda K \ (\lambda>0)$ to $T_{\rm G}$; that is, 
\begin{equation}
T_{\rm G}\approx \lambda \overline{K}^\prime+\sum_{s=1}^S \frac{r_{s}N}{Q_1(p=p_s)}.
\end{equation}
Conversely, we do not add that term to $T_{\rm R}$ because a random sequence means a sequence without sorting. 

We also define $\lambda_{\rm H}$ and $\lambda_{\rm M}$ as the values of $\lambda$ for which $T_{\rm H}=T_{\rm G}$ and $T_{\rm M}=T_{\rm G}$, respectively. Note that $\lambda_{\rm H}$ can have negative values, because $T_{\rm H}$ can be less than $T_{\rm G}$.

\renewcommand{\figurename}{TABLE}
\renewcommand{\thefigure}{\Roman{figure}}
\setcounter{figure}{5}
\renewcommand{\arraystretch}{1.6}
\begin{figure*}[htbp]
\centering
\caption{Upper and lower limits to $\lambda_{\rm cr}$}
\label{tab:lambda}
\begin{tabular}{c|c}
Region No. & Upper and lower limits to $\lambda_{\rm cr}$ \\ \hline \hline
1 & $0\leq \lambda<\displaystyle\frac{1}{r\alpha}\Biggl(\frac{p_1-\alpha^2}{p_1-\alpha}-\frac{p_2-\alpha^2}{p_2-\alpha}\Biggr)$ \\ 
2 & ${\rm max}\Biggl\{0,\displaystyle\frac{1}{r(1-r)}\Biggl(\frac{p_{\rm h}-\alpha^2}{\alpha(p_{\rm h}-\alpha)}-\frac{2r}{1-\sqrt{1-p_1}}-\frac{(1-r)(p_2-\alpha^2)}{\alpha(p_2-\alpha)}\Biggr)\Biggr\}\leq\lambda_{\rm cr}<\displaystyle\frac{1}{r}\Biggl(\frac{2}{1-\sqrt{1-p_1}}-\frac{p_2-\alpha^2}{\alpha(p_2-\alpha)}\Biggr)$\\ 
3 & ${\rm max}\Biggl\{0,\displaystyle\frac{1}{r(1-r)}\Biggl(\frac{2}{1-\sqrt{1-p_{\rm h}}}-\frac{2r}{1-\sqrt{1-p_1}}-\frac{(1-r)(p_2-\alpha^2)}{\alpha(p_2-\alpha)}\Biggr)\Biggr\}\leq\lambda_{\rm cr}<\displaystyle\frac{1}{r}\Biggl(\frac{2}{1-\sqrt{1-p_1}}-\frac{p_2-\alpha^2}{\alpha(p_2-\alpha)}\Biggr)$\\ 
4 & $\displaystyle\frac{2}{r(1-r)}\Biggl(\frac{1}{1-\sqrt{1-p_{\rm h}}}-\frac{r}{1-\sqrt{1-p_1}}-\frac{1-r}{1-\sqrt{1-p_2}}\Biggr)<\lambda_{\rm cr}<\displaystyle\frac{2}{r}\Biggl(\frac{1}{1-\sqrt{1-p_1}}-\frac{1}{1-\sqrt{1-p_2}}\Biggr)$
\end{tabular}
\end{figure*}
\renewcommand{\figurename}{FIG.}
\renewcommand{\thefigure}{\arabic{figure}}
\renewcommand{\arraystretch}{1}

\setcounter{figure}{16}
\begin{figure*}[htbp]
\centering
\includegraphics[width=17.5cm,clip]{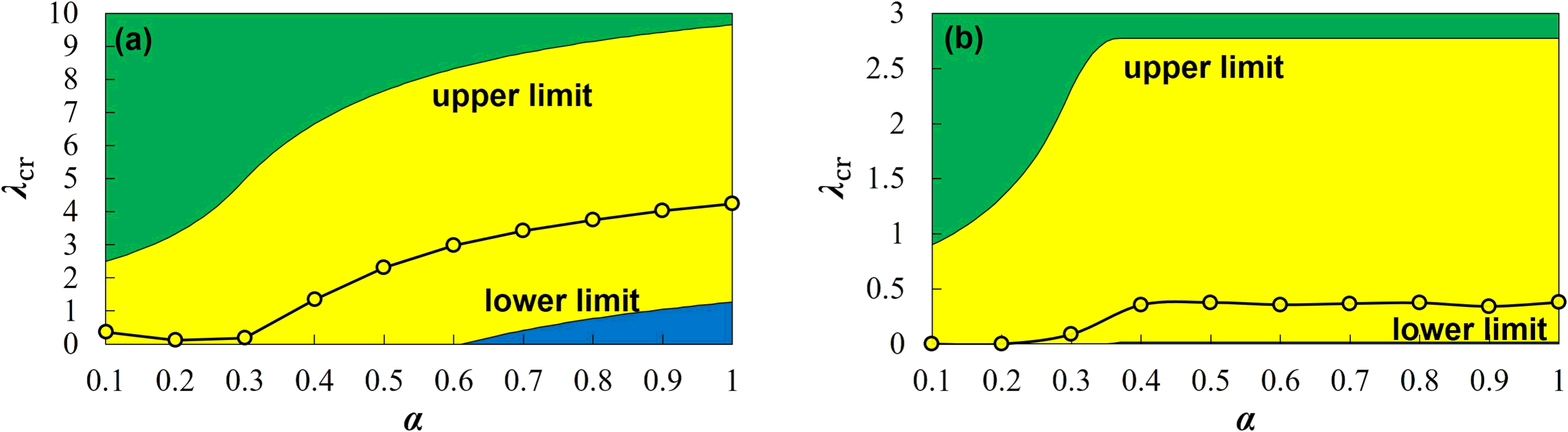}
\caption{(Color online) Simulation values (black circles) and the theoretical existence range of $\lambda_{\rm cr}$ (yellow region) as functions of $\alpha$. The other parameters are fixed at $(\beta;p_1,p_2;r)=$ (a) (1;0.5,1;0.5) and (b) (1;0.5,0.6;0.5).}
\label{fig:lambdaregion}
\end{figure*}

From Eq. (\ref{eq:TMTRTH}) and Subsec. \ref{sec:TRTG}, when $\lambda=0$ the relations among $T_{\rm R}$, $T_{\rm G}$, $T_{\rm H}$, and $T_{\rm M}$ must satisfy one of the following three inequalities:
\begin{equation}
T_{\rm H}<T_{\rm R}\leq T_{\rm G}<T_{\rm M},
\end{equation}
\begin{equation}
T_{\rm H}<T_{\rm G}< T_{\rm R}<T_{\rm M},
\end{equation}
or
\begin{equation}
T_{\rm G}<T_{\rm H}<T_{\rm R}<T_{\rm M}.
\end{equation}
Therefore, the relations of $\lambda_{\rm cr}$, $\lambda_{\rm H}$, and $\lambda_{\rm M}$ can be written as follows: 
\begin{equation}
{\rm max}(0, \lambda_{\rm H})\leq\lambda_{\rm cr}<\lambda_{\rm M},
\end{equation}
where we note that by definition $\lambda_{\rm cr}\geq0$, whereas $\lambda_{\rm H}$ can be either negative or positive, while $\lambda_{\rm M}$ must be positive.

In Region 1, i.e., $\alpha<1-\sqrt{1-p_1}$, where $T_{\rm H}<T_{\rm G}$ with $\lambda=0$, $\lambda_{\rm H}$ must be negative, while $\lambda_{\rm M}$ must be positive, and satisfy
\begin{equation}
\begin{split}
&\frac{N(p_1-\alpha^2)}{\alpha(p_1-\alpha)}\\
&\quad\approx\lambda_{\rm M} r(1-r)N +\frac{rN(p_1-\alpha^2)}{\alpha(p_1-\alpha)}+\frac{(1-r)N(p_2-\alpha^2)}{\alpha(p_2-\alpha)}.
\end{split}
\end{equation} 

In Region 2, i.e., $1-\sqrt{1-p_1}\leq\alpha<1-\sqrt{1-p_{\rm h}}$, $\lambda_{\rm H}$ can be either negative or positive, whereas $\lambda_{\rm M}$ must be positive. The quantities $\lambda_{\rm H}$ and $\lambda_{\rm M}$ satisfy 
\begin{equation}
\begin{split}
&\frac{rN(p_{\rm h}-\alpha^2)}{\alpha(p_{\rm h}-\alpha)}\\
&\quad\approx\lambda_{\rm H} r(1-r)N +\frac{2rN}{1-\sqrt{1-p_1}}+\frac{(1-r)N(p_2-\alpha^2)}{\alpha(p_2-\alpha)}
\end{split}
\end{equation}
and
\begin{equation}
\begin{split}
&\frac{2N}{1-\sqrt{1-p_1}}\\
&\quad\approx\lambda_{\rm M} r(1-r)N +\frac{2rN}{1-\sqrt{1-p_1}}+\frac{(1-r)N(p_2-\alpha^2)}{\alpha(p_2-\alpha)},
\end{split}
\end{equation} 
respectively. 

In Region 3, i.e., $1-\sqrt{1-p_{\rm h}}\leq\alpha<1-\sqrt{1-p_2}$, $\lambda_{\rm H}$ can be either negative or positive, and $\lambda_{\rm M}$ must be positive. Thus, $\lambda_{\rm H}$ and $\lambda_{\rm M}$ satisfy 
\begin{equation}
\begin{split}
&\frac{2N}{1-\sqrt{1-p_{\rm h}}}\\
&\quad\approx\lambda_{\rm H} r(1-r)N +\frac{2rN}{1-\sqrt1-p_1{}}+\frac{(1-r)N(p_2-\alpha^2)}{\alpha(p_2-\alpha)}
\end{split}
\end{equation}
and
\begin{equation}
\begin{split}
&\frac{2N}{1-\sqrt{1-p_1}}\\
&\quad\approx\lambda_{\rm M} r(1-r)N +\frac{2rN}{1-\sqrt{1-p_1}}+\frac{(1-r)N(p_2-\alpha^2)}{\alpha(p_2-\alpha)},
\end{split}
\end{equation} 
respectively. 

In Region 4, i.e., $1-\sqrt{1-p_2}\leq\alpha$, due to $T_{\rm G}<T_{\rm H}<T_{\rm M}$ when $\lambda=0$, $\lambda_{\rm H}$ and $\lambda_{\rm M}$ must both be positive. The quantities $\lambda_{\rm H}$ and $\lambda_{\rm M}$ therefore satisfy 
\begin{equation}
\begin{split}
&\frac{2N}{1-\sqrt{1-p_{\rm h}}}\\
&\quad\approx\lambda_{\rm H} r(1-r)N +\frac{2rN}{1-\sqrt{1-p_1}}+\frac{2(1-r)N}{1-\sqrt{1-p_2}}
\end{split}
\end{equation}
and
\begin{equation}
\begin{split}
&\frac{2N}{1-\sqrt{1-p_1}}\\
&\quad\approx\lambda_{\rm M} r(1-r)N +\frac{2rN}{1-\sqrt{1-p_1}}+\frac{2(1-r)N}{1-\sqrt{1-p_2}},
\end{split}
\end{equation} 
respectively.

Table \ref{tab:lambda} summarizes the upper and lower limits to $\lambda_{\rm cr}$ in each region. Furthermore, Fig. \ref{fig:lambdaregion} shows the simulation values (black circles) and the theoretical existence range of $\lambda_{\rm cr}$ (yellow region) as functions of $\alpha$. Note that the we calculated the simulation values using with 10-trial-averaged values of $T_{\rm R}$, $T_{\rm G}$, and $K$.

We can interpret Fig. \ref{fig:lambdaregion} as demonstrating that a group (random) sequence is preferable in the region below (above) the black line. In the blue (green) region, a group (random) sequence is in fact theoretically verified to be preferable. Comparing Figs. \ref{fig:lambdaregion} (a) and (b), the simulation values approach the lower limit---i.e., the accuracy of approximating $T_{\rm R}$ by $T_{\rm H}$ increases---as $|p_1-p_2|$ decreases. We admit that the yellow region is extensive, especially when $|p_1-p_2|$ is relatively large; however, we emphasize that the simulation values always exist within the expected region and that the region can be limited easily without numeric calculations, which is convenient for applications to actual situations.

\section{CONCLUSION}
\label{sec:conclusion}
In the present study, we have used a modified TASEP to analyze the dependence of the transportation time on the entering sequences of particles, using both the numerical simulations and theoretical analyses.

Here, we summarize a number of important results. In Sec. \ref{sec:simulation}, we discovered that there exists an important `group-advantageous region' where $T_{\rm R}> T_{\rm G}$ when ${\rm min}(\alpha,\beta)$ is relatively large and the sorting costs are neglected. 
When sorting costs are introduced,  a new region called a `random-advantageous region' appears with $T_{\rm R}< T_{\rm G}$. 
In addition, the group-advantageous region shrinks and finally disappears as $\lambda$ increases. We explored these phenomena for various $S\in\{2,3,N\}$.

In Sec. \ref{sec:theory}, we analyzed the simulation results by employing mathematical approaches for certain special cases. Using some approximations, we have shown theoretically that without the sorting cost the group-advantageous region must appear for any parameter sets ($S,p_s,r_s$). Moreover, we have succeeded in deriving the upper and lower limits to the value of $\lambda_{\rm cr}$ where $T_{\rm R}=T_{\rm G}$ by obtaining a general formula for the sorting cost.

Our findings can be applied to real-world situations, such as providing efficient operation for various tasks and smooth logistics for various products and yielding an effective evacuation method for pedestrians. Specifically, for smooth operation, we can determine whether we should begin tasks without considering the operation sequence or otherwise. Similarly, for smooth logistics, we can select whether the products should be bunched with nearly equal sizes. Furthermore, for ensuring effective evacuation of pedestrians, we can determine whether the bunching of pedestrians having nearly equal velocities should be conducted before transportation. The criteria for these judgments depend on the magnitude of the consideration or bunching cost ($\lambda$). Note that these magnitudes significantly differ from each other, i.e., considering only the sequence of tasks is typically deemed cheaper  (have a smaller $\lambda$) than sorting various pedestrians and products.
 
\section*{ACKNOWLEDGMENTS}
This work was partially supported by JST-Mirai Program Grant Number
JPMJMI17D4, Japan, JSPS KAKENHI Grant Number JP15K17583, and MEXT as
'Post-K Computer Exploratory Challenges' (Exploratory Challenge 2:
Construction of Models for Interaction Among Multiple Socioeconomic
Phenomena, Model Development and its Applications for Enabling Robust
and Optimized Social Transportation Systems) (Project ID: hp180188).

\appendix
\section{Validity of our selection of $L$ and $N$}
\label{sec:validity1}
In this Appendix, we briefly discuss the validity of selecting $L=200$ and $N=10,000$. 

\begin{figure}[htbp]
\begin{center}
\includegraphics[width=8cm,clip]{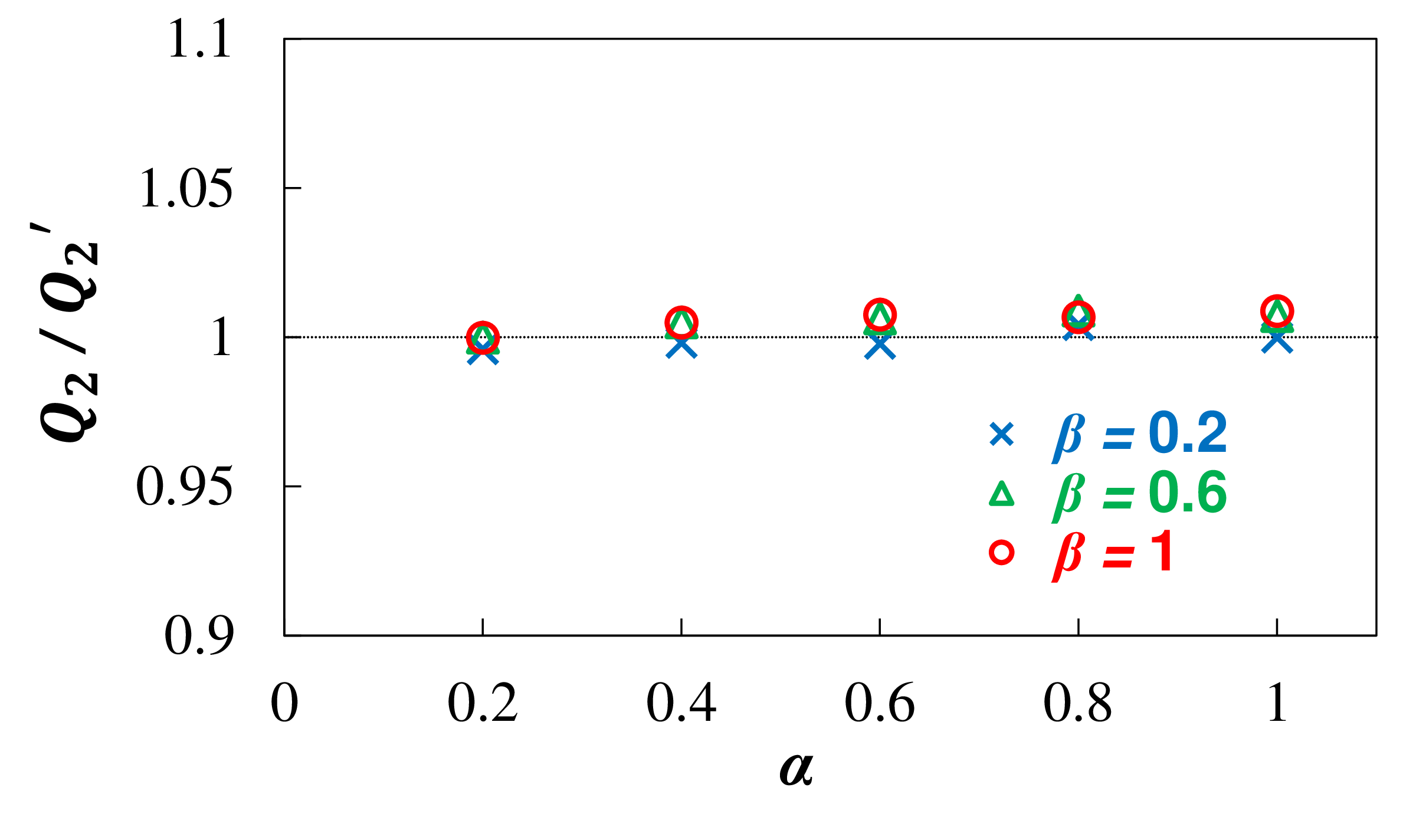}
\caption{(Color online) Simulation values of the ratio $Q_2/Q_2^{\prime}$ as a function of $\alpha$ for various $\beta \in \{0.2, 0.6, 1\}$. The other parameters are fixed at $(p_1,p_2;r)=(0.5,1;0.5)$.}
\label{fig:Ldepend}
\end{center}
\end{figure}

As finite-size effects may occur for small $L$, we compare the simulation values of $Q_2$ for $L=200$ and $L=1,000$. 
Figure \ref{fig:Ldepend} shows the ratio $Q_2/Q_2^{\prime}$, where $Q_2$ and $Q_2^{\prime}$ represent the flow of the multi-species TASEP with $L=200$ and $L=1,000$, respectively, as functions of $\alpha$ for various $\beta\in\{0.2,0.6,1\}$. The result that $Q_2/Q_2^{\prime}\approx 1$ indicates that the effect can be ignored for $L=200$. Thus, we choose $L=200$ to decrease the simulation time.

\begin{figure}[htbp]
\begin{center}
\includegraphics[width=8cm,clip]{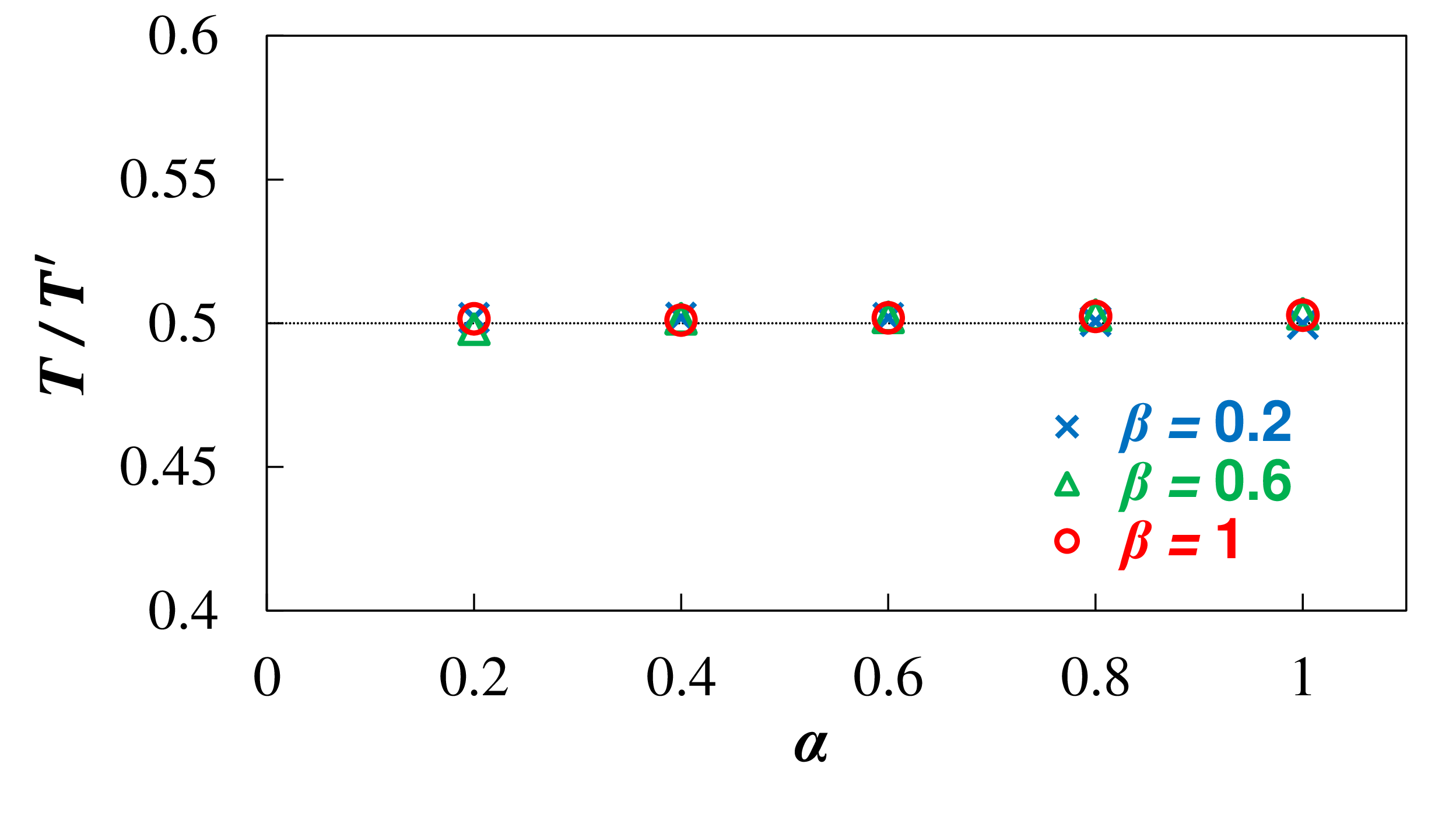}
\caption{(Color online) Simulation values of the ratio $T/T^{\prime}$ as a function of $\alpha$ for various $\beta \in \{0.2, 0.6, 1\}$. The other parameters are fixed at $L=200$ and $(p_1,p_2;r)=(0.5,1;0.5)$.}
\label{fig:Ndepend}
\end{center}
\end{figure}

On the other hand, the assumption that $T$ is determined by a steady-state flow may be inappropriate for small $N$. Therefore, we have compared the results for $N=10,000$ and $N=20,000$, in both cases for $L=200$. Figure \ref{fig:Ndepend} shows the ratio $T/T^{\prime}$, where $T$ and $T^{\prime}$ represent the transportation times for $N=10,000$ and $N=20,000$, respectively, as functions of $\alpha$ for various $\beta\in\{0.2,0.6,1\}$. The result that $T/T^{\prime}\approx 0.5$, i.e., that $T$ is proportional to $N$, indicates that the assumption can be regarded as valid for $N=10,000$. Thus, we choose $N=10,000$ similarly to decrease the simulation time.

\section{Simulation schemes for obtaining the minimal number of necessary exchanges}
\label{sec:sortalgo}
In this Appendix, we briefly describe the specific simulation schemes we used to obtain $K$. We emphasize that the cost of counting or comparing particles and the distances between exchanged particles are both ignored in the following.

First, for $S=2$, $\tau_{\rm a}=\tau_{\rm G}$ can have only one of two patterns. Once $\tau_{\rm G}$ is fixed to be either of these two sequences, we can immediately obtain the number of particles placed at the wrong areas in sequence $\tau_{\rm b}=\tau_{\rm R}$, which is twice as large as the number of necessary exchanges (see also Appendix \ref{sec:sort}). Consequently, comparing the results for the two $\tau_{\rm G}$ gives the smaller number as $K$. 

Second, for $S=3$, $\tau_{\rm a}=\tau_{\rm G}$ can have six patterns. Once $\tau_{\rm G}$ is fixed at one of these six sequences, we can immediately obtain the number of particles placed at the wrong areas in any sequence $\tau_{\rm b}=\tau_{\rm R}$. After selecting one species, which we first replace at the correct location, we exchange all particles of that species that are placed in the wrong areas in sequence $\tau_{\rm b}=\tau_{\rm R}$. The subsequent procedure is similar to the case for $S=2$. Consequently, comparing the six results for each $\tau_{\rm G}$ again gives the smallest number as $K$. Note that we can similarly calculate the numbers for general $S>4$.

Finally, for $S=N$, $\tau_{\rm a}=\tau_{\rm G}$ can have one of two patterns: either an ascending or a descending sequence. One exchange is needed for each particle in $\tau_{\rm b}=\tau_{\rm R}$ for which there exists a particle with a smaller (larger) hopping probability than the noted particle. This is termed a `selection sort.' This procedure starts from the leading particle. Consequently, by comparing the results for the two $\tau_{\rm G}$, the smaller number is again selected as the minimal number of necessary exchanges. 

\section{Probability distribution with $L=2$ and $S=2$}
\label{sec:probdis}
Here, we summarize the probability distributions with $L=2$ and $S=2$, which can be obtained from Eqs. (2) and (\ref{eq:normal}). The specific forms are described as follows:

\begin{equation}
\left\{
\begin{array}{l}
P_{00}=\displaystyle\frac{p_1p_2(1-\alpha)\beta^2}{\{(1-r)p_1+r p_2\}A+p_1p_2B}=\frac{p_{\rm h}(1-\alpha)\beta^2}{A+p_{\rm h}B}, \\ \\
P_{0\ast}=\displaystyle\frac{p_1p_2\alpha\beta}{\{(1-r)p_1+r p_2\}A+p_1p_2B}=\frac{p_{\rm h}\alpha\beta}{A+p_{\rm h}B}, \\ \\
P_{10}=\displaystyle\frac{rp_2 A}{\{(1-r)p_1+r p_2\}A+p_1p_2B}=\frac{rp_{\rm h}A}{p_1(A+p_{\rm h}B)}, \\ \\
P_{1\ast}=\displaystyle\frac{rp_1p_2\alpha^2(1-\beta)}{\{(1-r)p_1+r p_2\}A+p_1p_2B}=\frac{rp_{\rm h}\alpha^2(1-\beta)}{A+p_{\rm h}B}, \\ \\
P_{20}=\displaystyle\frac{(1-r)p_1 A}{\{(1-r)p_1+r p_2\}A+p_1p_2B}=\frac{(1-r)p_{\rm h}A}{p_2(A+p_{\rm h}B)}, \\ \\
\begin{split}
P_{2\ast}&=\displaystyle\frac{(1-r)p_1p_2\alpha^2(1-\beta)}{\{(1-r)p_1+r p_2\}A+p_1p_2B} \\ \\
&=\frac{(1-r)p_{\rm h}\alpha^2(1-\beta)}{A+p_{\rm h}B},
\end{split}
\end{array}
\right.
\end{equation}
where
\begin{equation}
A=\alpha\beta(\alpha+\beta-\alpha\beta),
\end{equation}
\begin{equation}
B=\alpha^2+\beta^2-\alpha^2\beta-\alpha\beta^2+\alpha\beta,
\end{equation}
and
\begin{equation}
p_{\rm h}=\frac{p_1 p_2}{(1-r)p_1+r p_2}.
\end{equation}

\section{$Q_S$ for general $S$ with $L=2$}
\label{sec:generals}
In this appendix, we prove that for general $S$ with $L=2$, $Q_S$ is equal to $Q_1(p=p_{\rm h})$.

From the results with $L=2$ and $S=2$ (see Appendix \ref{sec:probdis}), we can conjecture the probability distributions for general $S$ with $L=2$ as

\begin{equation}
\left\{
\begin{array}{l}
P_{00}=\displaystyle\frac{p_{\rm h}(1-\alpha)\beta^2}{A+p_{\rm h}B}, \\ \\
P_{0\ast}=\displaystyle\frac{p_{\rm h}\alpha\beta}{A+p_{\rm h}B}, \\ \\
P_{s0}=\displaystyle\frac{r_s p_{\rm h}A}{p_s(A+p_{\rm h}B)}, \\ \\
P_{s\ast}=\displaystyle\frac{r_s p_{\rm h}\alpha^2(1-\beta)}{A+p_{\rm h}B},
\end{array}
\right.
\label{eq:expect}
\end{equation}
where $s=1,2,......,S$.

On the other hand, the master equations of the steady state are summarized as $2(S+1)$ equations:

\begin{equation}
\left\{
\begin{array}{l}
P_{00}=(1-\alpha)P_{00}+(1-\alpha)\beta P_{0\ast}, \\ \\
P_{0\ast}=(1-\alpha)(1-\beta)P_{0\ast}+\displaystyle\sum_{k=1}^{S}p_k P_{k0}, \\ \\
P_{s0}=r_s P_{00}+r_s\alpha\beta P_{0\ast}+(1-p_s) P_{s0}+\beta P_{s\ast}, \\ \\
P_{s\ast}=(1-\beta)P_{s\ast}+r_s \alpha(1-\beta)P_{0\ast},
\end{array}
\right.
\label{eq:master2}
\end{equation}
where $s=1,2,......,S$.
In addition, $P_{ij}$ must satisfy  the normalization condition

\begin{equation}
\sum_{i=0}^S  P_{i0} + \sum_{i=0}^S  P_{i\ast}= 1.
\label{eq:normal2}
\end{equation}

We can confirm that Eqs. (\ref{eq:expect}) satisfy Eqs. (\ref{eq:master2}) and (\ref{eq:normal2}). With Penron-Frobenius theorem regarding stochastic matrix, this indicates that Eqs. (\ref{eq:expect}) are unique solutions for Eqs. (\ref{eq:master2}) and (\ref{eq:normal2}).

From Eqs. (\ref{eq:expect}), the flow of the system is given by the following expression:

\begin{equation}
\begin{split}
Q_S&=\sum_{s=1}^{S}p_s P_{s0}=\displaystyle\sum_{s=1}^{S}p_s \frac{r_s p_{\rm h}A}{p_s(A+p_{\rm h}B)} \\
&=Q_1(p=p_{\rm h}).
\end{split}
\label{eq:QS}
\end{equation}

\section{Validity of the approximation for $T$}
\label{sec:validity2}
In this Appendix, we briefly demonstrate the validity of Eq. (\ref{eq:Tg}). 

\begin{figure}[htbp]
\begin{center}
\includegraphics[width=8cm,clip]{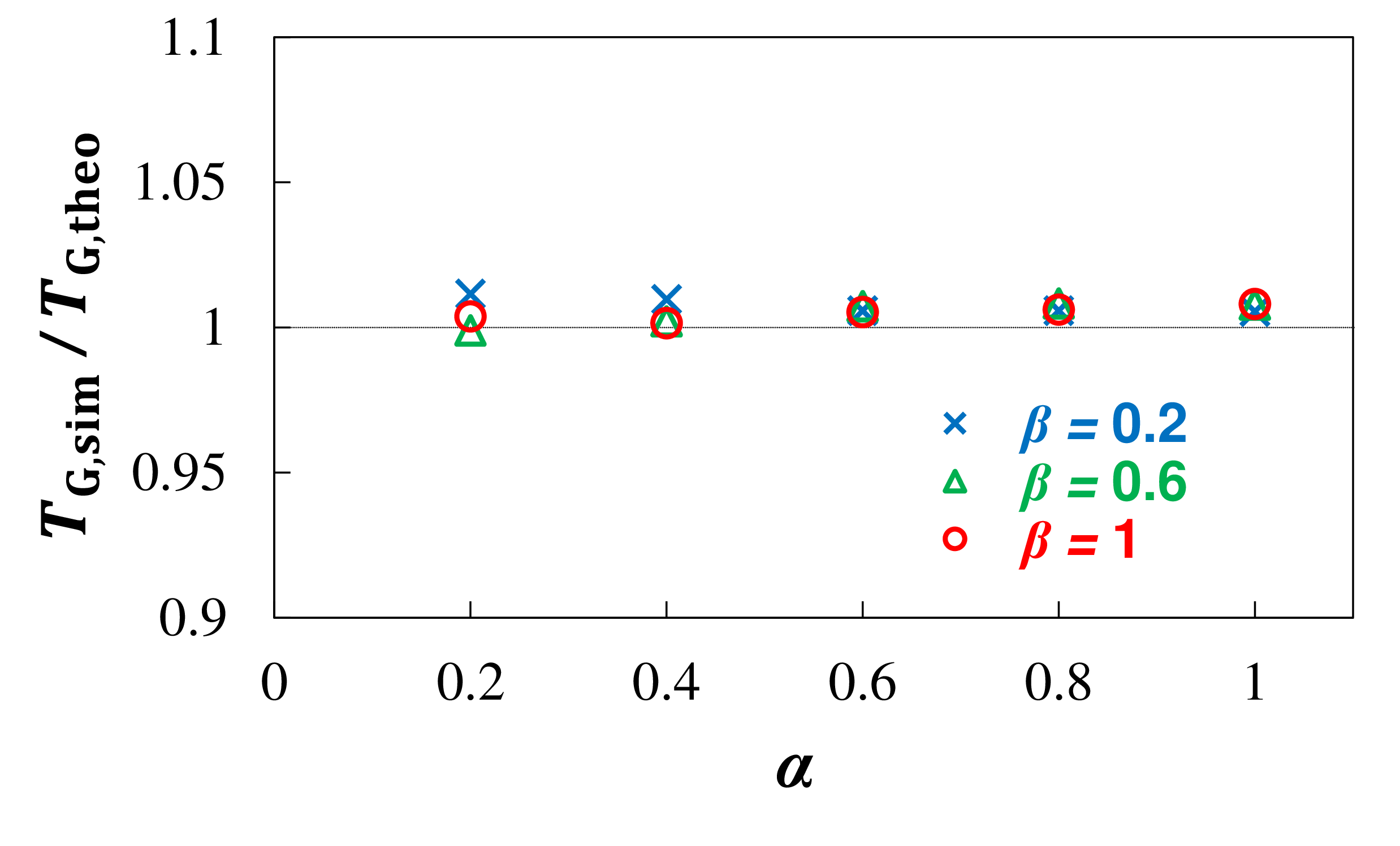}\\
\includegraphics[width=8cm,clip]{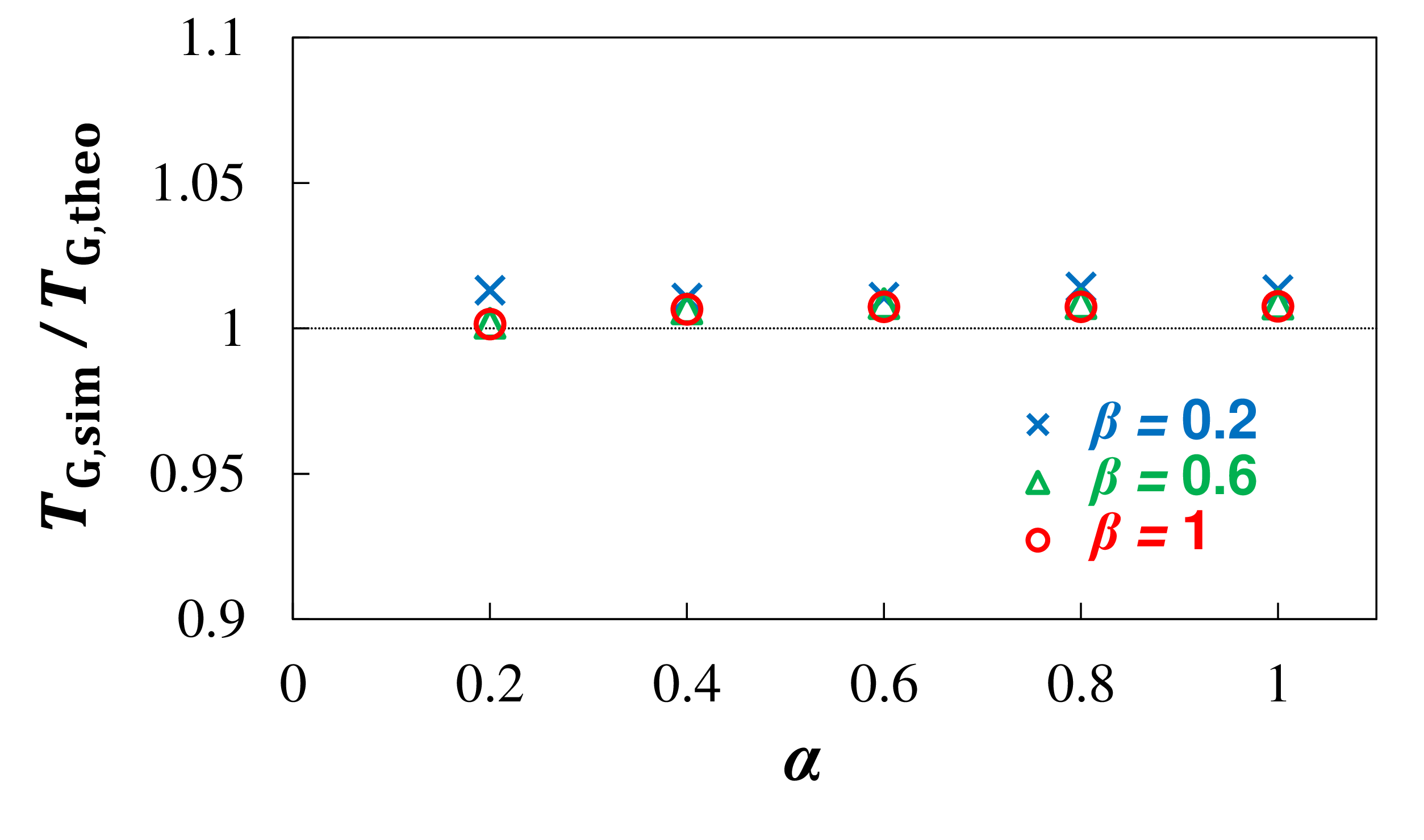}
\caption{(Color online) Simulations values of the ratio $T_{\rm G,sim}/T_{\rm G,theo}$ as a function of $\alpha$ for various $\beta \in \{0.2, 0.6, 1\}$ with (a) $S=2$ and (b) $S=3$. The other parameters are fixed at (a) $(p_1,p_2;r)=(0.5,1;0.5)$ and (b) $(p_1,p_2,p_3;r_1,r_2,r_3)=(0.4,0.6,0.8;0.2,0.3,0.5)$.}
\label{fig:validityT1}
\end{center}
\end{figure}

Figure \ref{fig:validityT1} (a) shows the ratio $T_{\rm G,sim}/T_{\rm G,theo}$ as a function of $\alpha$ for various $\beta\in\{0.2,0.6,1\}$ with $S=2$ and Fig. \ref{fig:validityT1} (b) shows the same ratio for $S=3$. Note that $T_{\rm G,sim}$ and $T_{\rm G,theo}$ represent the values of $T_{\rm G}$ from the simulations and that given by Eq. (\ref{eq:Tg}), respectively. Both figures show that $T_{\rm G,sim}/T_{\rm G,theo}\approx1$, indicating that Eq. (\ref{eq:Tg}) provides a good approximation for $T_{\rm G}$. 

Strictly speaking, $T_{\rm G,sim}/T_{\rm G,theo}$ must be larger than 1 on average. This is mainly due to the fact that $T_{\rm G,sim}$ includes $T_1$, which is the time required for the first particle to reach the right-hand boundary, whereas $T_{\rm G,theo}$ ignores that time. This also indicates that $T_{\rm G}$ can differ depending on the order of each group in the group sequence (i.e., the hopping probability of the leading group). However, this difference has little influences on the theoretical results, as explained below. 

\begin{figure}[htbp]
\centering
\includegraphics[width=8cm,clip]{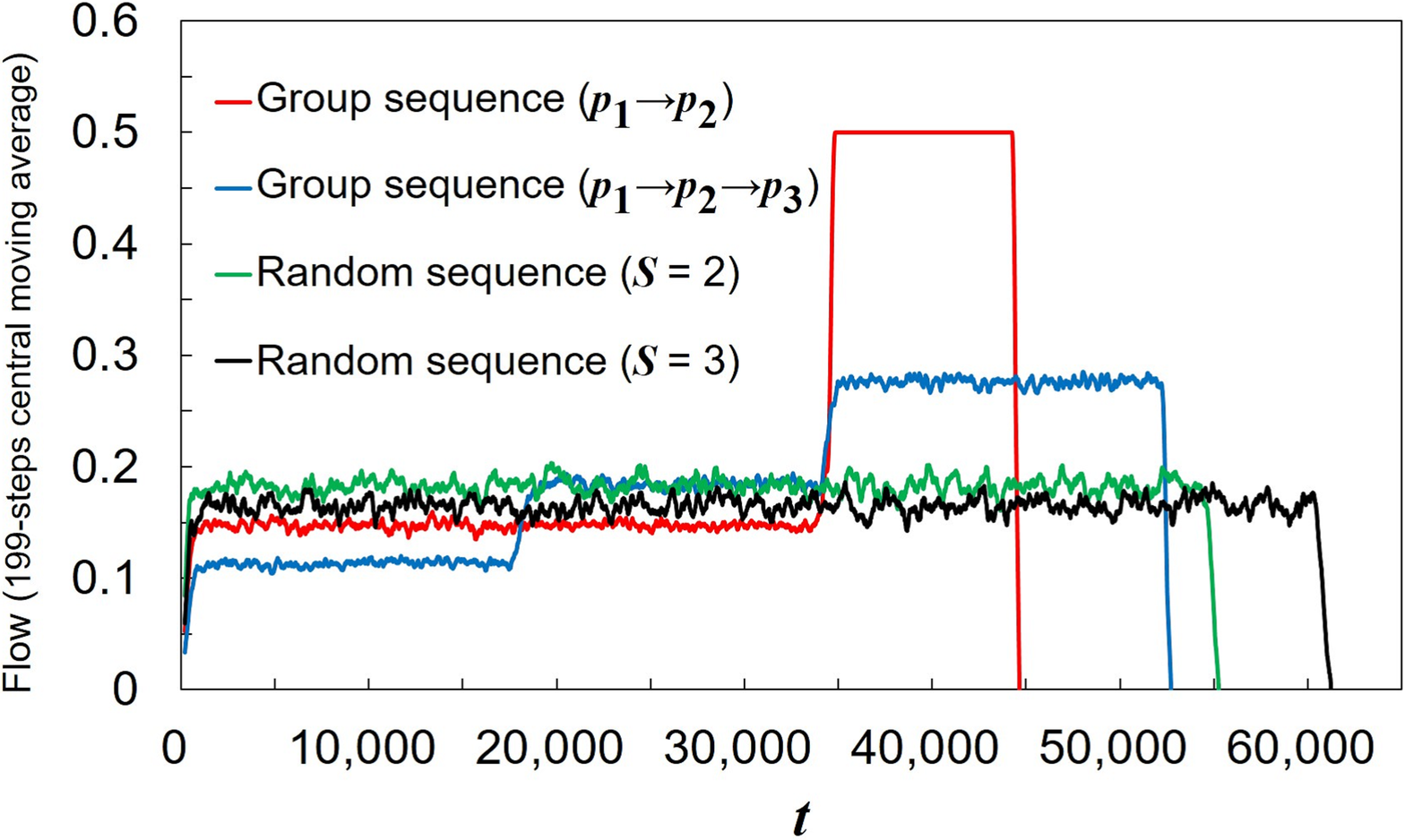}
\caption{(Color online) Simulation values of 199-steps central moving average of flow at time $t$ with $(N,\alpha,\beta)=(10,000,1,1)$. For $S=2$ and $S=3$, respectively, we set $(p_1,p_2;r)=(0.5,1;0.5)$, $(p_1,p_2,p_3;r_1,r_2,r_3)=(0.4,0.6,0.8;0.2,0.3,0.5)$. Note that each plot ends just after it reaches 0.}
\label{fig:flow}
\end{figure}

First, $T_1$ can be estimated as
\begin{equation}
T_1\approx\frac{L}{p_s},
\end{equation}
where $s=1,2,......,S$ and the time steps before the first particle enters the lattice are assumed to be small enough to be ignored. The quantities $T_1$ and $T_G$ without $T_1$ satisfy
\begin{equation}
T_1\approx\frac{L}{p_s}<\frac{L}{p_S}<\frac{L}{1-\sqrt{1-p_S}}
\end{equation}
and
\begin{equation}
T_{\rm G}\approx\sum_{s=1}^{S} \frac{r_s N}{Q_1(p=p_s)}>\frac{2N}{1-\sqrt{1-p_S}},
\end{equation}
respectively. Therefore, $T_1/T_{\rm G}$ reduces to 
\begin{equation}
\frac{T_1}{T_{\rm G}}<\frac{L}{2N}.
\end{equation}
Under the proposition that $N$ is large enough, we can assume $L/2N<<1$ ($L/2N=0.01$ in the present study). In fact, observing the time series of the flows (199-steps central moving average) in Fig. \ref{fig:flow}, we find that nearly the entire duration during transportation can be regarded to be in the steady state for large enough $r_sN$. Note that we calculate the flows at time $t$ by averaging number of moving particles per bond between $t-1$ and $t$. Moreover, all the transportation times $T$ ($T_{\rm R}$, $T_{\rm H}$, and $T_{\rm M}$) originally include $T_1$, so that this term disappears when they are subtracted from each other. Consequently, $T_1$ (and therefore, the dependence of $T_{\rm G}$ on the order of each group in the group sequence) can be assumed to be ignorable.

\section{Discussion of the sign of $f(\alpha)$ \\ in Region 1}
\label{sec:region1app}
In this Appendix, we give a detailed derivation of Eq. (\ref{eq:fLD}) for Region 1, where $\alpha<1-\sqrt{1-p_1}$.

Eq. (\ref{eq:fregion1}) gives $f(\alpha)$ 
\begin{equation}
\begin{split}
f(\alpha)&=\frac{N(p_{\rm h}-\alpha^2)}{\alpha(p_{\rm h}-\alpha)}-\sum_{s=1}^{S} \frac{r_s N(p_s-\alpha^2)}{\alpha(p_s-\alpha)}\\
&=\frac{N}{\alpha}\sum_{s=1}^{S} \left\{\frac{r_s (p_{\rm h}-\alpha^2)}{p_{\rm h}-\alpha}-\frac{r_s (p_s-\alpha^2)}{p_s-\alpha}\right\}\\
&=\displaystyle\frac{N(\alpha-1)}{(p_{\rm h}-\alpha)\prod_{s=1}^{S}(p_s -\alpha)}C,
\end{split}
\label{eq:differld}
\end{equation}
where 
\begin{equation}
C=\sum_{s=1}^{S} \left\{r_s(p_{\rm h}-p_s)\prod_{k\neq s}(p_k -\alpha)\right\}.
\end{equation}
The quantity $C$ is calculated as follows:
\begin{equation}
\begin{split}
C&=\sum_{s=1}^{S}\left\{r_s\left(\displaystyle\frac{1}{\sum_{t=1}^{S}r_t/p_t}-p_s\right)\prod_{k\neq s}(p_k -\alpha)\right\}\\
&=\frac{1}{D}\sum_{s=1}^{S}\left\{r_s\left(\prod_{k=1}^{S}p_k-p_s \sum_{t=1}^{S}r_t\prod_{k\neq t}p_k\right)\prod_{k\neq s}(p_k -\alpha)\right\}\\
&=\frac{1}{D}\sum_{s=1}^{S}\sum_{t=1}^{S}\left\{r_s r_t\left(\prod_{k=1}^{S}p_k-p_s\prod_{k\neq t}p_k\right)\prod_{k\neq s}(p_k -\alpha)\right\}\\
&=\frac{1}{D}\sum_{s=1}^{S}\sum_{t\neq s}\left\{r_s r_t(p_t-p_s)\prod_{k\neq t}p_k\prod_{k\neq s}(p_k -\alpha)\right\},
\end{split}
\label{eq:C}
\end{equation}
where 
\begin{equation}
D=\sum_{s=1}^{S}\left(r_s\prod_{k\neq s}p_k\right).
\label{eq:D}
\end{equation}
By regarding the sum of the term with $(s,t)=(x,y)$ and that with $(s,t)=(y,x)$ as a new term for $\exists (x, y) \ (x,y=1,2,......,S, x<y)$, we can rewrite Eq. (\ref{eq:C}) as follows:
\begin{equation}
\begin{split}
C&=\frac{1}{D}\sum_{s=1}^{S}\sum_{t<s}\Biggl\{r_s r_t(p_t-p_s)\prod_{k\neq t}p_k\prod_{k\neq s}(p_k -\alpha)\\
&\quad\quad\quad\quad\quad\quad+r_t r_s(p_s-p_t)\prod_{k\neq s}p_k\prod_{k\neq t}(p_k -\alpha)\Biggr\}\\
&=\frac{1}{D}\sum_{s=1}^{S}\sum_{t<s}\left\{\alpha r_s r_t(p_t-p_s)^2\prod_{k\neq s, t}p_k\prod_{k\neq s, t}(p_k -\alpha)\right\}.
\end{split}
\label{eq:D2}
\end{equation}
Because $p_s -\alpha>0$ ($\forall s$) and $p_{\rm h} -\alpha>0$, we obtain $C>0$.

Considering $\alpha-1<0$ and $C>0$, we finally obtain    
\begin{equation}
f(\alpha)<0.
\end{equation}

\section{Continuity and differentiability of $f(\alpha)$ at each boundary}
\label{sec:differentiable}
In this Appendix, we briefly discuss the continuity and differentiability of $f(\alpha)$ at each boundary. 

Defining $g(x)$ for $0<x\leq1$ as
\begin{eqnarray}
g(x)=\left\{ \begin{array}{ll}
\displaystyle  \frac{p-x^2}{x(p-x)} & {\rm for} \ 0<x\leq 1-\sqrt{1-p}, \\
\displaystyle\frac{2}{1-\sqrt{1-p}} & {\rm for} \ 1-\sqrt{1-p}<x\leq1,
\end{array} \right.
\label{eq:gx}
\end{eqnarray}
where $0<p\leq1$, the following equations hold:
\begin{equation}
\lim_{x \to q-0} g(x)=\lim_{x \to q+0} g(x)=\frac{2}{1-\sqrt{1-p}}
\end{equation}
and
\begin{equation}
\lim_{\delta \to -0} \frac{g(x+\delta)-g(x)}{\delta}=\lim_{\delta \to +0} \frac{g(x+\delta)-g(x)}{\delta}=0,
\end{equation}
where $q=1-\sqrt{1-p}$. Therefore, $g(x)$ is continuous and differentiable at $x=q=1-\sqrt{1-p}$, resulting in the continuity and differentiability of $g(x)$ for $0<x\leq1$.

As a result, because $f(\alpha)$ is represented as a linear sum of terms $g(\alpha)$, where $p$ is substituted for $p_s$ or $p_{\rm h}$ ($0<p_s,p_{\rm h}\leq0$), $f(\alpha)$ is clearly continuous and differentiable at each boundary.

\section{Discussion of the sign of $df(\alpha)/d\alpha$ \\ in Subregion 2--$v$}
\label{sec:region2app}
In this Appendix, we discuss the sign of $df(\alpha)/d\alpha$ in Subregion 2--$v$, i.e., $1-\sqrt{1-p_{v}}\leq\alpha<1-\sqrt{1-p_{v+1}}$

From Eq. (\ref{eq:fregion2}), $df(\alpha)/d\alpha$ can be calculated as follows:

\begin{equation}
\begin{split}
&\frac{d f(\alpha)}{d\alpha}\\
&\approx \frac{N(2\alpha-\alpha^2-p_{\rm h})}{\alpha^2(p_{\rm h}-\alpha)^2}-\sum_{s=u}^{S} \frac{r_sN(2\alpha-\alpha^2-p_s)}{\alpha^2(p_{\rm s}-\alpha)^2}\\
&= \sum_{s=1}^{v} \frac{r_sN(2\alpha-\alpha^2-p_{\rm h})}{\alpha^2(p_{\rm h}-\alpha)^2}\\
&\quad+\sum_{s=v+1}^{S} \Biggl\{\frac{r_sN(2\alpha-\alpha^2-p_{\rm h})}{\alpha^2(p_{\rm h}-\alpha)^2}-\frac{r_sN(2\alpha-\alpha^2-p_s)}{\alpha^2(p_{\rm s}-\alpha)^2}\Biggr\}\\
&= \sum_{s=1}^{v} \frac{r_sN(2\alpha-\alpha^2-p_{\rm h})}{\alpha^2(p_{\rm h}-\alpha)^2}\\
&\quad+\sum_{s=v+1}^{S} \frac{r_sN(p_s-p_{\rm h})\{(\alpha-1)^2+p_s+p_{\rm h}-p_sp_{\rm h}-1\}}{p_{\rm h}p_s(p_{\rm h}-\alpha)^2(p_{\rm s}-\alpha)^2}.
\end{split}
\label{eq:fprimeregion2app}
\end{equation}
For $1-\sqrt{1-p_v}\leq\alpha<1-\sqrt{1-p_{v+1}}$, the following two inequalities hold:
\begin{equation}
\centering
\begin{split}
&2\alpha-\alpha^2-p_{\rm h}\\
&<2(1-\sqrt{1-p_{v+1}})-(1-\sqrt{1-p_{v+1}})^2-p_{\rm h}\\
&=p_{v+1}-p_{\rm h}<0
\end{split}
\label{eq:region2app1}
\end{equation}
and
\begin{equation}
\centering
\begin{split}
&(\alpha-1)^2+p_s+p_{\rm h}-p_sp_{\rm h}-1\\
&>(1-\sqrt{1-p_{v+1}}-1)^2+p_s+p_{\rm h}-p_sp_{\rm h}-1\\
&>(1-\sqrt{1-p_{\rm h}}-1)^2+p_s+p_{\rm h}-p_sp_{\rm h}-1\\
&=p_s(1-p_{\rm h})>0.
\end{split}
\label{eq:region2app2}
\end{equation}

We cannot specify the sign of $df(\alpha)/d\alpha$ in this subregion from Eqs. (\ref{eq:fprimeregion2app}), (\ref{eq:region2app1}), and (\ref{eq:region2app2}). However, near the boundary between Subregion 2--$(u-1)$ and 3--$u$, we obtain the following conditions:
\begin{equation}
\begin{split}
&\lim_{\alpha\to q_h-0} (2\alpha-\alpha^2-p_{\rm h})\\
&= 2(1-\sqrt{1-p_{\rm h}})-(1-\sqrt{1-p_{\rm h}})^2-p_{\rm h}=0
\end{split}
\end{equation}
and
\begin{equation}
\begin{split}
&\lim_{\alpha\to q_h-0} \{(\alpha-1)^2+p_s+p_{\rm h}-p_sp_{\rm h}-1\}\\
&= (1-\sqrt{1-p_{\rm h}}-1)^2+p_{u-1}+p_{\rm h}-p_{u-1}p_{\rm h}-1\\
&=p_{u-1}(1-p_{\rm h})>0,
\end{split}
\end{equation}
where $q_h=1-\sqrt{1-p_{\rm h}}$.
Therefore, noting the obvious continuity of $df(\alpha)/d\alpha$ for $1-\sqrt{1-p_{v}}\leq\alpha<1-\sqrt{1-p_{v+1}}$, the region of $df(\alpha)/d\alpha>0$ must exist at least in Subregion 2--$(u-1)$.

\section{Discussion of the sign of $df(\alpha)/d\alpha$ \\ in Subregion 3--$v$}
\label{sec:region3app}
In this Appendix, we give a proof on Eq. (\ref{eq:fprimeregion3}) in Subregion 3--$v$, i.e., $1-\sqrt{1-p_{v-1}}\leq\alpha<1-\sqrt{1-p_{v}}$.

From Eq. (\ref{eq:fregion3}), $d f(\alpha)/d\alpha$ can be calculated as follows:

\begin{equation}
\frac{d f(\alpha)}{d\alpha}\approx \sum_{s=v}^{S} \frac{r_sN(p_s-2\alpha+\alpha^2)}{\alpha^2(p_{\rm s}-\alpha)^2}.
\label{eq:fprimeregion3app}
\end{equation}
For $s=v,......,S$, the quantity $p_s-2\alpha+\alpha^2$ satisfies 
\begin{equation}
\begin{split}
&p_s-2\alpha+\alpha^2\\
&>p_s-2(1-\sqrt{1-p_{v}})+(1-\sqrt{1-p_{v}})^2\\
&=p_s-p_{v}>0.
\end{split}
\label{eq:region3app1}
\end{equation}
From Eqs. (\ref{eq:fprimeregion3app}) and (\ref{eq:region3app1}), we finally obtain 
\begin{equation}
\frac{d f(\alpha)}{d\alpha}>0.
\end{equation}

\section{Discussion of the sign of $f(\alpha)$ \\ in Region 4}
\label{sec:region4app}
In this Appendix, we give a detailed derivation of Eq. (\ref{eq:fMC}), where $1-\sqrt{1-p_S}\leq \alpha$.

From Eqs. (\ref{eq:Tg2}) and (\ref{eq:Th2}), $f(\alpha)$ can be represented as follows:
\begin{equation}
\begin{split}
T_{\rm G}&\approx 2N\sum_{s=1}^{S} \frac{r_s (1+\sqrt{1-p_s})}{p_s}\\
&=\frac{2N}{\prod_{s=1}^{S}p_s}\left\{\sum_{s=1}^{S} \left(r_s+r_s\sqrt{1-p_s}\right) {\prod_{t\neq s} p_t}\right\}\\
&=\frac{2N}{\prod_{s=1}^{S} p_s}\Biggl\{ \sum_{s=1}^{S} \left(r_s\prod_{t\neq s} p_t\right)\\
&\quad\quad\quad\quad\quad+\sum_{s=1}^{S} \left(r_s\sqrt{1-p_s}\prod_{t\neq s} p_t\right) \Biggr\}
\end{split}
\label{eq:Tg4}
\end{equation}
and
\begin{equation}
\begin{split}
T_{\rm H}&\approx 2N\frac{1+\sqrt{1-p_{\rm h}}}{p_{\rm h}}\\
&=2N \frac{1+\sqrt{1-1/\sum_{s=1}^{S} (r_s/p_s)}}{1/\sum_{s=1}^{S}(r_s/p_s)}\\
&=\frac{2N}{\prod_{s=1}^{S} p_s} \Biggl\{ \sum_{s=1}^{S} \left(r_s\prod_{t\neq s} p_t \right)\\
&\quad+\sqrt{\left(\sum_{s=1}^{S}r_s\prod_{t\neq s} p_t\right)^2-\prod_{s=1}^{S}p_s\times \left(\sum_{s=1}^{S}r_s \prod_{t\neq s} p_t\right)} \Biggr\}
\end{split}
\label{eq:Th4}
\end{equation}
From Eqs. (\ref{eq:fregion4}), (\ref{eq:Tg4}) and (\ref{eq:Th4}), $f(\alpha)$ is given by

\begin{equation}
\begin{split}
&f(\alpha)\\
&\approx\displaystyle\frac{2N}{\prod_{s=1}^{S}p_s} \Biggl\{\sqrt{\Biggl(\sum_{s=1}^{S}r_s\prod_{t\neq s} p_t\Biggr)^2-\prod_{s=1}^{S}p_s\times \Biggl(\sum_{s=1}^{S}r_s \prod_{t\neq s} p_t}\Biggr)\\
&\quad-\sum_{s=1}^{S}\Biggl(r_s\sqrt{1-p_s}\prod_{t\neq s} p_t\Biggr)\Biggr\}\\ 
&=\frac{2N}{\prod_{s=1}^{S}p_s}(E-F),
\end{split}
\label{eq:differ}
\end{equation}
where 
\begin{equation}
E=\sqrt{\left(\sum_{s=1}^{S}r_s\prod_{t\neq s} p_t\right)^2-\prod_{s=1}^{S}p_s\times\left(\sum_{s=1}^{S}r_s \prod_{t\neq s} p_t\right)}
\label{eq:E}
\end{equation}
and
\begin{equation}
F=\sum_{s=1}^{S}\left(r_s\sqrt{1-p_s}\prod_{t\neq s} p_t\right).
\label{eq:F}
\end{equation}

From Eqs. (\ref{eq:E}) and (\ref{eq:F}), $E^2-F^2$ becomes
\begin{equation}
\begin{split}
&E^2-F^2\\
&=\left(\sum_{s=1}^{S}r_s\prod_{t\neq s} p_t\right)^2-\prod_{s=1}^{S}p_s\times \left(\sum_{s=1}^{S}r_s \prod_{t\neq s} p_t\right)\\
&\quad-\left\{\sum_{s=1}^{S}\left(r_s\sqrt{1-p_s}\prod_{t\neq s} p_t\right)\right\}^2\\ 
&=\sum_{s=1}^{S}\left(r_s\prod_{t\neq s} p_t\right)^2-\sum_{s=1}^{S}\sum_{t\neq s}\left\{r_s r_t\prod_{k\neq s} p_k\prod_{l\neq t} p_l\right\} \\
&\quad-\prod_{s=1}^{S}p_s\times \left(\sum_{s=1}^{S}r_s \prod_{t\neq s} p_t\right)\\
&\quad-\sum_{s=1}^{S}\left(r_s\prod_{t\neq s} p_t\right)^2+\sum_{s=1}^{S}\left\{r_s^2p_s\left(\prod_{t\neq s} p_t\right)^2\right\}\\
&\quad-\sum_{s=1}^{S}\sum_{t\neq s}\left\{r_s r_t\sqrt{1-p_s}\sqrt{1-p_t}\prod_{k\neq s} p_k\prod_{l\neq t} p_l\right\}\\
&=\sum_{s=1}^{S}\Biggl\{\sum_{t\neq s}r_s r_t\prod_{k\neq s}p_k\prod_{l\neq t}p_l\\
&\quad\quad\quad-r_s(1-r_s)p_t\prod_{k\neq s}p_k\prod_{l\neq t}p_l\\
&\quad\quad\quad-\sum_{t\neq s}\Biggl(r_s r_t\sqrt{1-p_s}\sqrt{1-p_t}\prod_{k\neq s} p_k\prod_{l\neq t} p_l\Biggr)\Biggr\}\\
&=\sum_{s=1}^{S}\sum_{t\neq s}\Biggl[r_sr_t\prod_{k\neq s}p_k\prod_{l\neq t}p_l\times\\
&\quad\quad\quad\quad\quad\Biggl\{1-p_t-\sqrt{(1-p_s)(1-p_t)}\Biggr\}\Biggr].
\end{split}
\label{eq:differ2}
\end{equation}
Here, regarding the sum of the term with $(s,t)=(x,y)$ and that with $(s,t)=(y,x)$ as a new term for $\exists (x, y) \ (x,y=1,2,......,S, x<y)$, Eq. (\ref{eq:differ2}) can be rewritten as
\begin{equation}
\begin{split}
&E^2-F^2\\
&=\sum_{s=1}^{S}\sum_{t<s}\Biggl[r_sr_t\prod_{k\neq s}p_k\prod_{l\neq t}p_l\times\\
&\quad\quad\quad\quad\quad\Biggl\{2-p_s-p_t-\sqrt{(1-p_s)(1-p_t)}\Biggr\}\Biggr]\\
&=\sum_{s=1}^{S}\sum_{t<s}\left\{r_sr_t\left(\sqrt{1-p_s}-\sqrt{1-p_t}\right)^2\prod_{k\neq s}p_k\prod_{l\neq t}p_l\right\}.
\end{split}
\label{eq:differ3}
\end{equation}

Due to Eq. (\ref{eq:differ3}) and the non-negativity of both $E$ and $F$, we have $E> F$, thereby resulting in 
\begin{equation}
f(\alpha)>0.
\end{equation}

\section{Specific conditions on $\alpha_{\rm cr, max}$}
\label{sec:f0}
Here, we discuss the specific conditions on $\alpha_{\rm cr, max}$. 

Table \ref{tab:alphacr} summarizes explicit expressions for $f(\alpha=\alpha_{\rm cr})=0$, where the upper (lower) expression holds in Region 2 (Region 3). For $S=2$, the lower expression becomes a quadratic equation in $\alpha_{\rm cr}$. However, the upper expression becomes a quartic equation that is too difficult to solve analytically those conditions. Note that for $S>2$, both equations become more than quartic.

From its definition of $\alpha_{\rm cr,max}$, $\alpha_{\rm cr,max}$ can be written as
\begin{equation}
\alpha_{\rm cr,max}={\rm max}\{\alpha_{\rm cr}\},
\end{equation}
where $\{\alpha_{\rm cr}\}$ represents the set of $\alpha_{\rm cr}$.   

\renewcommand{\figurename}{TABLE}
\renewcommand{\thefigure}{\Roman{figure}}
\setcounter{figure}{5}
\begin{figure*}[htbp]
\begin{center}
\renewcommand{\arraystretch}{1.6}
\centering
\caption{Explicit expressions for $f(\alpha=\alpha_{\rm cr}$)=0}
\label{tab:alphacr}
\begin{tabular}{c|c}
Region No. & Explicit expressions for $f(\alpha=\alpha_{\rm cr})=0$ \\ \hline \hline
2 & \begin{tabular}{c}
$\displaystyle\frac{N(p_{\rm h}-\alpha_{\rm cr}^2)}{\alpha_{\rm cr}(p_{\rm h}-\alpha_{\rm cr})}-\displaystyle\sum_{s=1}^{v} \frac{2r_s N}{1-\sqrt{1-p_s}}-\displaystyle\sum_{s=v+1}^{S} \frac{r_s N(p_s-\alpha_{\rm cr}^2)}{\alpha_{\rm cr}(p_s-\alpha_{\rm cr})}=0$ 
\\ \quad\quad  $\land$ \ $\left\{\begin{array}{ll}
1-\sqrt{1-p_{v}}<\alpha_{\rm cr}<1-\sqrt{1-p_{v+1}} & {\rm for} \ 1\leq v\leq u-2  \\
1-\sqrt{1-p_{v}}<\alpha_{\rm cr}<1-\sqrt{1-p_{\rm h}} & {\rm for} \ v=u-1
\end{array}\right.$  
\end{tabular}\\ \hline
3& \begin{tabular}{c}
$\displaystyle\frac{2N}{1-\sqrt{1-p_{\rm h}}}-\displaystyle\sum_{s=1}^{v-1} \frac{2r_s N}{1-\sqrt{1-p_s}}-\sum_{s=v}^{S} \frac{r_s N(p_s-\alpha_{\rm cr}^2)}{\alpha_{\rm cr}(p_s-\alpha_{\rm cr})}=0$ 
\\ \quad\quad $\land$ \ $\left\{\begin{array}{ll}
1-\sqrt{1-p_{\rm h}}<\alpha_{\rm cr}<1-\sqrt{1-p_{v}} & {\rm for} \ v=u  \\
1-\sqrt{1-p_{v-1}}<\alpha_{\rm cr}<1-\sqrt{1-p_{v}} & {\rm for} \ u+1\leq v\leq S
\end{array}\right.$ 
\end{tabular} 
\end{tabular}
\end{center}
\end{figure*}
\renewcommand{\figurename}{FIG.}
\renewcommand{\thefigure}{\arabic{figure}}
\renewcommand{\arraystretch}{1}

\section{Derivation of $\overline{K}^\prime$}
\label{sec:sort}
In this Appendix, we derive the approximate averaged minimal number of exchanges $\overline{K}^\prime$ necessary to sort the particles for two special cases: $S=2$ and $S=N$. 

\subsection{$S=2$}
First, for a general calculation of $\overline{K}^\prime$, $\tau_{\rm G}$ has to be fixed to be either of the two possible patterns. Once $\tau_{\rm G}$ is fixed, $K(\tau_{\rm G},\tau_{\rm R})$ can be determined uniquely for all possible $\tau_{\rm R}$. Without loss of generality, we can assume $r N\leq (1-r) N$ and $\tau_{\rm G}$ can be fixed as illustrated in the lower panel of Fig. \ref{fig:K}.

\setcounter{figure}{21}
\begin{figure}[htbp]
\begin{center}
\includegraphics[width=8.5cm,clip]{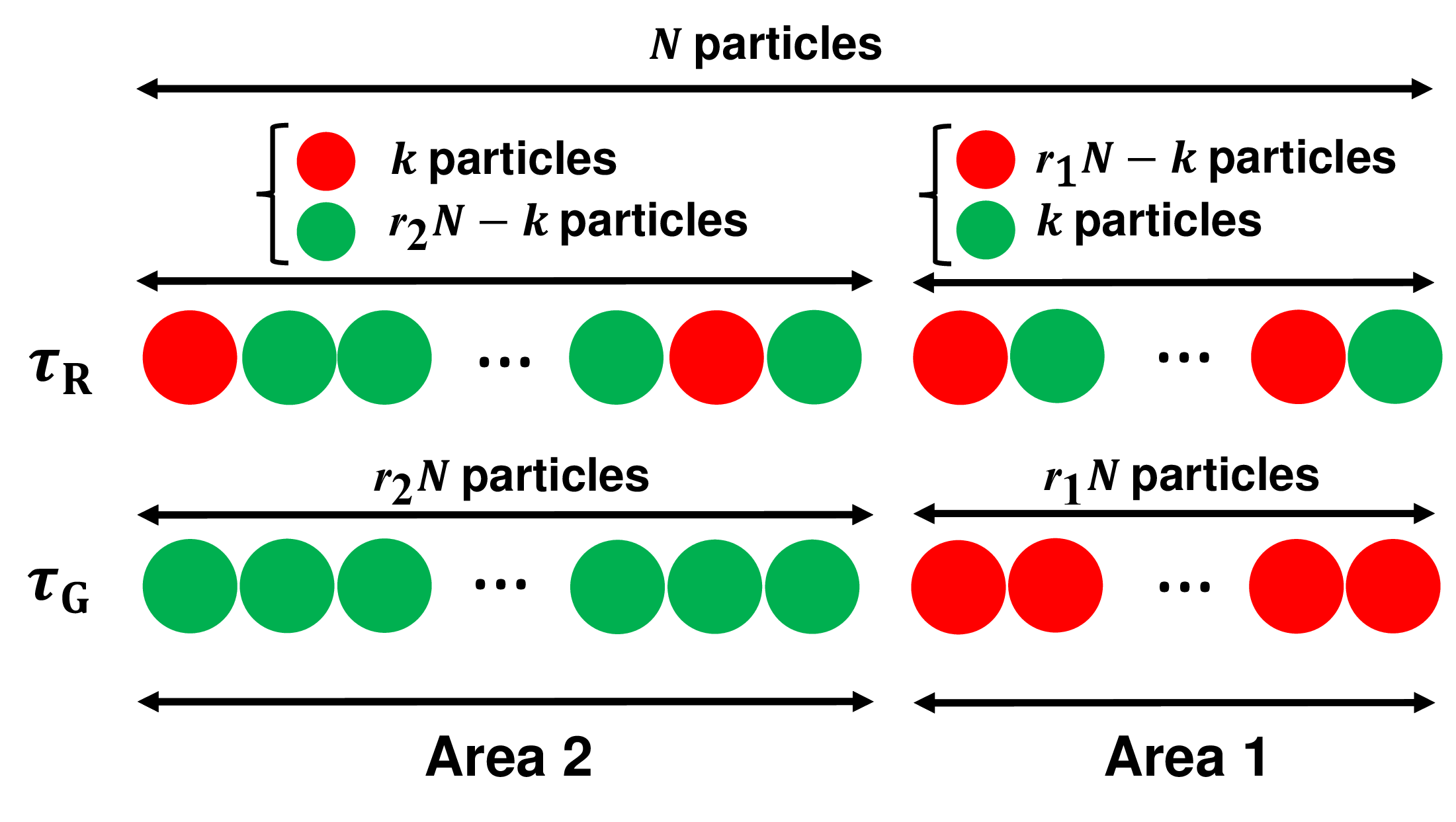}
\caption{(Color online) Schematic illustration of $\tau_{\rm G}$ (upper panel) and $\tau_{\rm R}$ (lower panel), where the red particles belong to species 1 and the green ones to species 2. In the upper panel, we show one example from among all $\binom{r N}{k}\times\binom{(1-r) N}{k}$ possible random sequences, whereas in the lower panel we show one of the two possible group sequences.}
\label{fig:K}
\end{center}
\end{figure}

Suppose that for $\tau_{\rm R}$, $k \ (0\leq k \leq r N)$ particles of species 1 are located in the Area 2, ($k$ particles of species 2 are located in the Area 1, conversely) as described in the lower of Fig. \ref{fig:K}. Under this supposition, $k$-time exchanges are necessary for sorting particles from $\tau_{\rm R}$ to $\tau_{\rm G}$. Considering that $\tau_{\rm R}$ satisfying this supposition possibly has $\binom{r N}{k}\times \binom{(1-r) N}{k}$ sequences, $a_N=\sum_{\forall{\tau_{\rm R}}}K^{\prime}(\tau_{\rm G},\tau_{\rm R})$ can be written as follows;

\begin{equation}
\begin{split}
a_N&=\sum_{\forall{\tau_{\rm R}}}K^{\prime}(\tau_{\rm G},\tau_{\rm R})\\
&=\sum_{k=1}^{r N} k\binom{r N}{k}\binom{(1-r) N}{k}\\
&=\sum_{k=1}^{r N} r N\binom{r N-1}{k-1}\binom{(1-r) N}{k}\\
&=r N\sum_{k=1}^{r N} \Biggl\{\binom{r N}{k}\binom{(1-r) N}{k}\\
&\quad\quad\quad\quad\quad-\binom{r N-1}{k}\binom{(1-r) N}{k}\Biggr\}.
\end{split}
\label{eq:KN}
\end{equation}
Using the Vandermonde convolution formula, Eq. (\ref{eq:KN}) can be rewritten as follows:
\begin{equation}
a_N=rN\left\{\binom{N}{rN}-\binom{N-1}{rN}\right\}.
\label{eq:KN2}
\end{equation}
Because the sequence $\tau_{\rm R}$ can take any of $N!/\{(r N)!((1-r) N)!\}$ possible patterns with equal probability, we can finally reduce $\overline{K}^\prime$ to
\begin{equation}
\overline{K}^\prime=\frac{(r N)!((1-r) N)!}{N!}a_N=r(1-r) N.
\label{eq:KN3}
\end{equation}

\begin{figure}[htbp]
\centering
\includegraphics[width=8cm,clip]{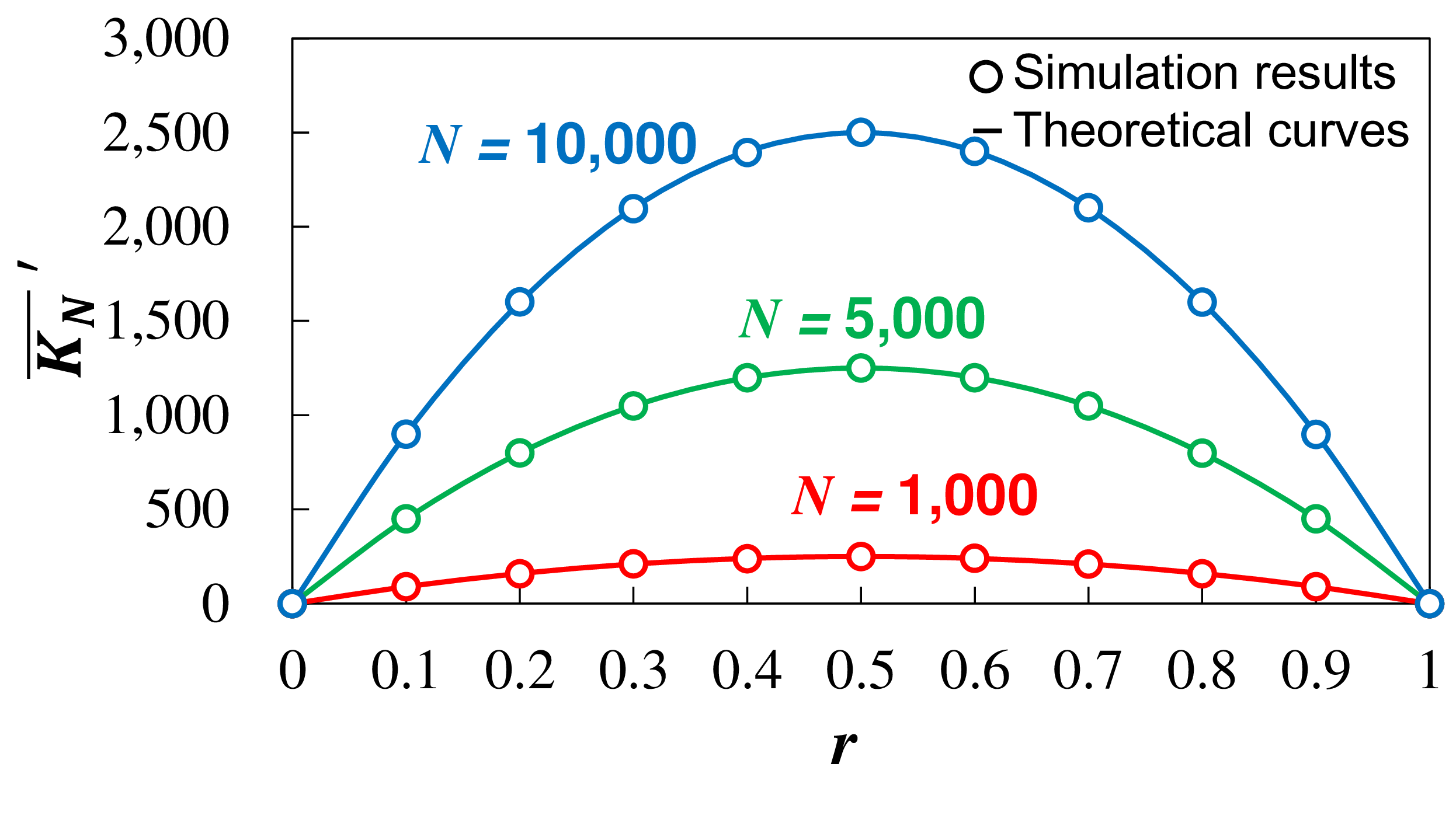}
\caption{(Color online) Simulation (circles) and theoretical (curve) values of $\overline{K}^\prime$  as functions of $r$ for various $N\in\{1,000({\rm red}), 5,000({\rm green}), 10,000({\rm blue})\}$ with $S=2$. We obtained each of the simulation values by averaging over 100 trials.}
\label{fig:sortS2}
\end{figure}

Figure \ref{fig:sortS2} compares the simulation (circles) and theoretical (curves) values for various $N\in\{1,000({\rm red}), 5,000({\rm green}), 10,000({\rm blue})\}$ for $S=2$. The simulations show very good agreement with our exact analysis.

\subsection{$S=N$}
\begin{figure}[htbp]
\centering
\includegraphics[width=8cm,clip]{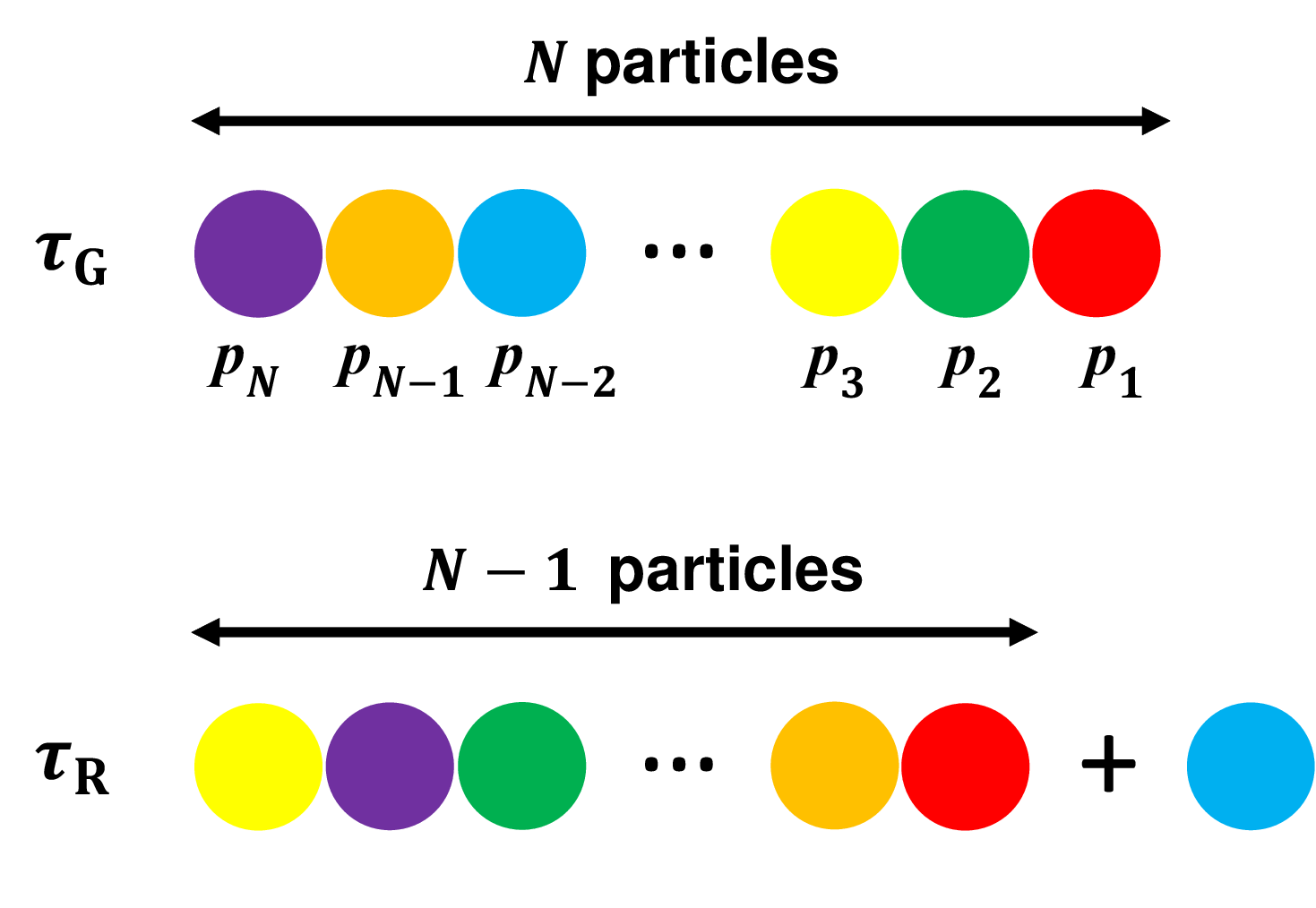}
\caption{(Color online) Schematic illustration of $\tau_{\rm G}$ (upper panel) and $\tau_{\rm G}$ (lower panel) for the case $S=N$. In the lower panel, we show one example of all $N\times (N-1)!$ possible sequences. Note that $p_1<p_2<...<p_N$ for the ascending sequence, whereas $p_1>p_2>...>p_N$ for the descending one.}
\label{fig:KN}
\end{figure}

When $S=N$, $\tau_{\rm G}$ also has to be fixed as either of the two possible patterns---an ascending or a descending sequence---for a general calculation of $\overline{K}^\prime$, as illustrated in the upper panel of Fig. \ref{fig:KN}. Once $\tau_{\rm G}$ is fixed, $K^{\prime}(\tau_{\rm G},\tau_{\rm R})$ can be determined uniquely for all possible $\tau_{\rm R}$. 

If we regard the entire sequence as consisting of two parts---the first (blue) particle and other ($N-1$) particles, as described in the lower panel of Fig. \ref{fig:KN}---the sorting procedure can also be divided into two parts: sorting ($N-1$) particles plus the last exchange for the first particle. If the first particle corresponds to the particle with hopping probability $p_l$ $(l=1,2,......,N)$, and noting that the sequence for the remaining ($N-1$) particles has $(N-1)!$ possible patterns, we can calculate the quantity $b_{N,l}=\sum_{\forall{\tau_{{\rm R},l}^\prime}}K(\tau_{\rm G}, \tau_{{\rm R},l}^\prime)$ as follows:

\begin{eqnarray}
\begin{split}
b_{N,l}&=\sum_{\forall{\tau_{{\rm R},l}^\prime}}K(\tau_{\rm G}, \tau_{{\rm R},l}^\prime)\\
&=\left\{ \begin{array}{ll}
a_{N-1} & {\rm for} \ l=1, \\
a_{N-1}+(N-1)! & {\rm for} \ l=2,3,......,N,
\end{array} \right.
\end{split}
\end{eqnarray}
where $N>1$ and $\tau_{{\rm R},l}^\prime$ represents the sequence for which the first particle is the particle with hopping probability $p_l$. Note that the last sort is not necessary in the case where $l=1$.

Therefore, for $N>1$, we can write $a_N=\sum_{\forall{\tau_{\rm R}}}K(\tau_{\rm G},\tau_{\rm R})$:
\begin{equation}
\begin{split}
a_N&=\sum_{\forall{\tau_{\rm R}}}K_N(\tau_{\rm G},\tau_{\rm R})\\
&=\sum_{l=1}^N\sum_{\forall{\tau_{{\rm R},l}^\prime}}K_N(\tau_{\rm G}, \tau_{{\rm R},l}^\prime)\\
&=\sum_{l=1}^N b_{N,l}\\
&=(N-1)\times(N-1)!+N a_{N-1}
\end{split}
\label{eq:an}
\end{equation}
Dividing both sides of Eq. (\ref{eq:an}) by $N!$, we obtain
\begin{equation}
c_N=c_{N-1}+\frac{N-1}{N}=c_1+\sum_{k=1}^N\frac{k-1}{k},
\end{equation}
where $c_N=a_N/N!$ and $N>2$. With the initial condition $c_1=a_1=0$, $a_N$ is finally reduced to
\begin{equation}
a_N=N!\left(N-\sum_{k=1}^N \frac{1}{k}\right),
\end{equation}
which we note holds for the case $N=1$.

The sequence $\tau_{\rm R}$ can take $N!$ patterns with equal probability, and therefore, $\overline{K}^\prime$ is finally reduced to
\begin{equation}
\overline{K}^\prime=\frac{a_N}{N!}=N-\sum_{k=1}^N \frac{1}{k}.
\end{equation}

\begin{figure}[htbp]
\centering
\includegraphics[width=8cm,clip]{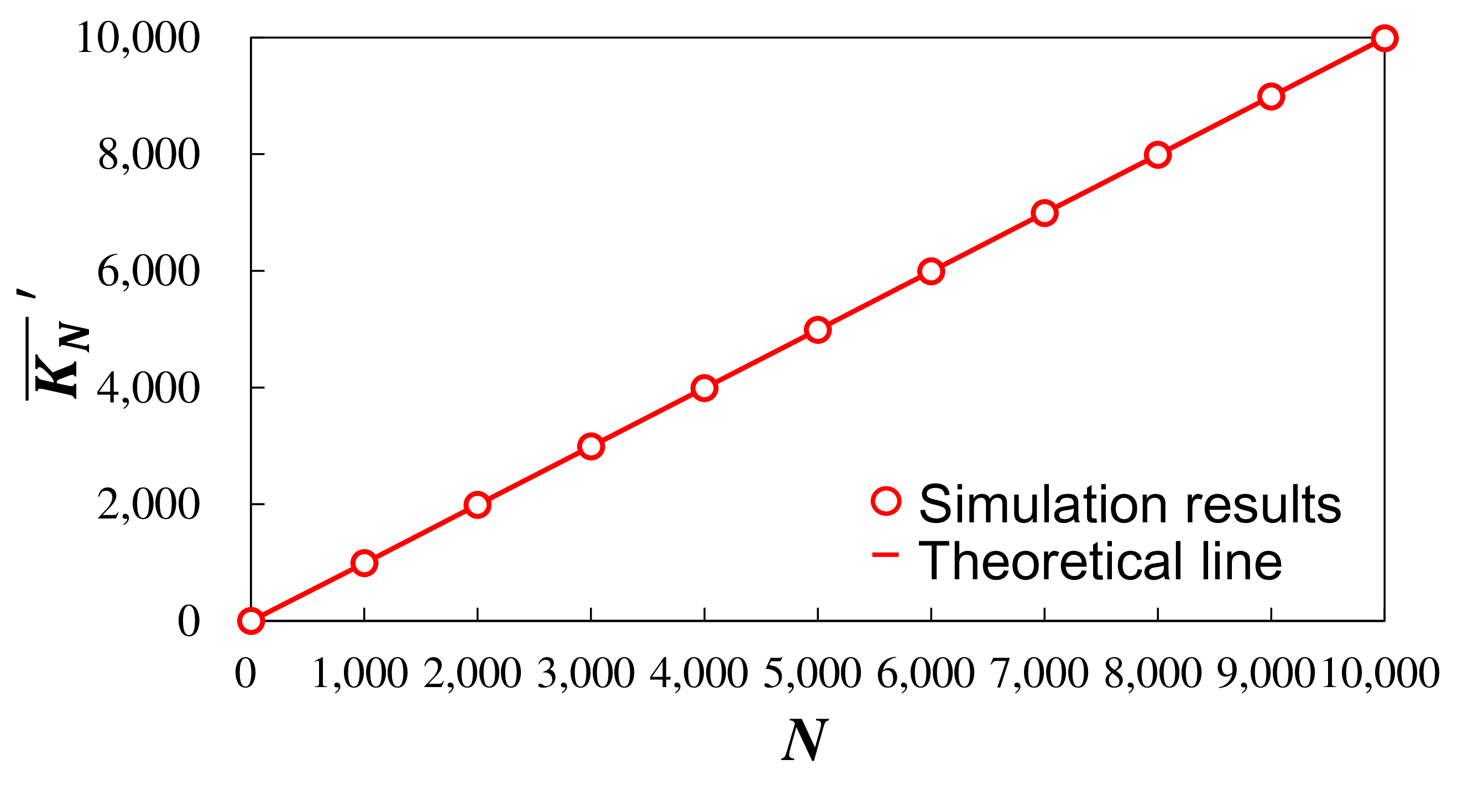}
\caption{(Color online) Simulation (circles) and theoretical (line) values of $\overline{K_N}^\prime$ as functions of $N$. We obtained each of the simulation values by averaging over 100 trials.}
\label{fig:sortSN}
\end{figure}

Figure \ref{fig:sortSN} compares the simulation (circles) and theoretical (line) values for $S=N$. The simulations again show a very good agreement with our exact analysis.

\clearpage

\end{document}